\documentclass[
reprint, 
superscriptaddress,
amsmath,
amssymb,
aps,
prb,
longbibliography,
]{revtex4-2}

\usepackage{graphicx}
\usepackage{dcolumn}
\usepackage{physics}
\usepackage{gensymb}
\usepackage{bm}
\usepackage{xcolor}
\usepackage{footmisc}
\usepackage{siunitx}
\usepackage{tabularx}
\usepackage{comment}
\usepackage{hyperref}

\usepackage[normalem]{ulem}

\begin{document}

\preprint{APS/123-QED}


\title{A physically motivated analytical expression for the temperature dependence of the zero-field splitting of the nitrogen-vacancy center in diamond}

\author{M.~C.~Cambria}
 
\affiliation{Department of Physics, University of Wisconsin, Madison, WI 53706, USA}

\author{G.~Thiering}

\affiliation{Wigner Research Centre for Physics, P.O. Box 49, 1525 Budapest, Hungary}

\author{A.~Norambuena}
 
\affiliation{Centro de Optica e Informaci\'on Cu\'antica, Universidad Mayor, Camino la Piramide 5750, Huechuraba, Santiago, Chile}

\author{H.~T.~Dinani}
 
\affiliation{Escuela de Ingenier\'{i}a, Facultad de Ciencias, Ingenier\'{i}a y Tecnolog\'{i}a, Universidad Mayor, Santiago, Chile}

\author{A.~Gardill}

\author{I.~Kemeny}
 
\affiliation{Department of Physics, University of Wisconsin, Madison, WI 53706, USA}

\author{V.~Lordi}
 
\affiliation{Lawrence Livermore National Laboratory, Livermore, CA, 94551, USA}

\author{\'A.~Gali}

\affiliation{Wigner Research Centre for Physics, P.O. Box 49, 1525 Budapest, Hungary}

\affiliation{Department of Atomic Physics, Institute of Physics, Budapest University of Technology and Economics, M\H{u}egyetem rakpart 3., 1111 Budapest, Hungary}

\author{J.~R.~Maze}
 
\affiliation{Instituto de F\'isica, Pontificia Universidad Cat\'olica de Chile, Casilla 306, Santiago, Chile}

\affiliation{Centro de Investigaci\'on en Nanotecnolog\'ia y Materiales Avanzados, Pontificia Universidad Cat\'olica de Chile, Santiago, Chile}

\author{S.~Kolkowitz}
\email{kolkowitz@berkeley.edu}
 
\affiliation{Department of Physics, University of Wisconsin, Madison, WI 53706, USA}

\affiliation{Department of Physics, University of California, Berkeley, CA 94720, USA
}%

\date{\today}
\begin{abstract}
The temperature dependence of the zero-field splitting (ZFS) between the $|m_{s}=0\rangle$ and $|m_{s}=\pm 1\rangle$ levels of the nitrogen-vacancy (NV) center's electronic ground-state spin triplet can be used as a robust nanoscale thermometer in a broad range of environments. However, despite numerous measurements of this dependence in different temperature ranges, to our knowledge no analytical expression has been put forward that captures the scaling of the ZFS of the NV center across all relevant temperatures. Here we present a simple, analytical, and physically motivated expression for the temperature dependence of the NV center's ZFS that matches all experimental observations, in which the ZFS shifts in proportion to the occupation numbers of two representative phonon modes. In contrast to prior models our expression does not diverge outside the regions of fitting. We show that our model quantitatively matches experimental measurements of the ZFS from 15 to 500 K in single NV centers in ultra-pure bulk diamond, and we compare our model and measurements to prior models and experimental data.
\end{abstract}

\maketitle

\textit{Introduction.}---
The zero-field splitting (ZFS) quantifies the energy difference between a spin system's \(m_{s}\) levels in the absence of externally applied fields. This value plays a critical role in determining the properties of spin defects in crystals, which have emerged as leading platforms for quantum sensing \cite{abendroth2022single, qiu2022nanoscale, chen2023real, rovny2022nanoscale} and quantum networking \cite{bradley2022robust, stas2022robust}. Besides characterizing a spin defect's level structure, the ZFS can also be temperature dependent and thereby provides a mechanism for nanoscale thermometry. This effect has already been exploited with the nitrogen-vacancy (NV) center in diamond for \emph{in vivo} thermometry \cite{kucsko2013nanometre, fujiwara2020real, choi2020probing} and thermal conductivity measurements \cite{laraoui2015imaging}. In the context of the NV center, the ZFS refers to the splitting between the \(m_{s} = 0\) and \(m_{s}=\pm1\) levels of the negatively charged NV center's electronic ground-state spin triplet. 

The centrality of the ZFS to NV physics and its applications to thermometry has motivated a number of prior works examining its temperature dependence \cite{acosta2010temperature, chen2011temperature, toyli2012measurement, doherty2014temperature, li2017temperature, barson2019temperature, lourette2022temperature, tang2022first}. While various analytical expressions have been fit to the measured NV ZFS temperature dependence over specific temperature ranges, to our knowledge all prior models make use of power series expansions or other approximate polynomial expressions, and as such diverge outside the range of experimental data used to fit the models and tend to provide little physical insight \cite{chen2011temperature, toyli2012measurement, li2017temperature, barson2019temperature}. Recent \textit{ab initio} efforts have demonstrated near quantitative agreement between numerical calculations and experiment, but lack useful analytical expressions \cite{tang2022first}. These factors motivate the development of a model of the NV ZFS temperature dependence that is both analytical and physically motivated, such that the model is predictive and practically useful over a wide temperature range, and also provides insight into the spin-lattice interactions that give rise to the temperature dependence.

In this work, we measure the ZFS as a function of temperature in multiple, as-grown, single NV centers in high-purity bulk diamond over a wide range of experimentally relevant temperatures, from 15 to 500 K. We present a novel, simple analytical model of the ZFS temperature dependence in which the ZFS shifts in proportion to the occupation numbers of two representative phonon modes. We provide physical justification for the model by demonstrating that it describes the effects of both first- and second-order atomic displacements. Finally, we discuss the advantages of the proposed model over other currently available models and we highlight the applicability of the proposed model to other spin-lattice effects in spin defects. 

\begin{figure}[t]
\vspace{-0.3cm}
\includegraphics[width=0.48\textwidth]{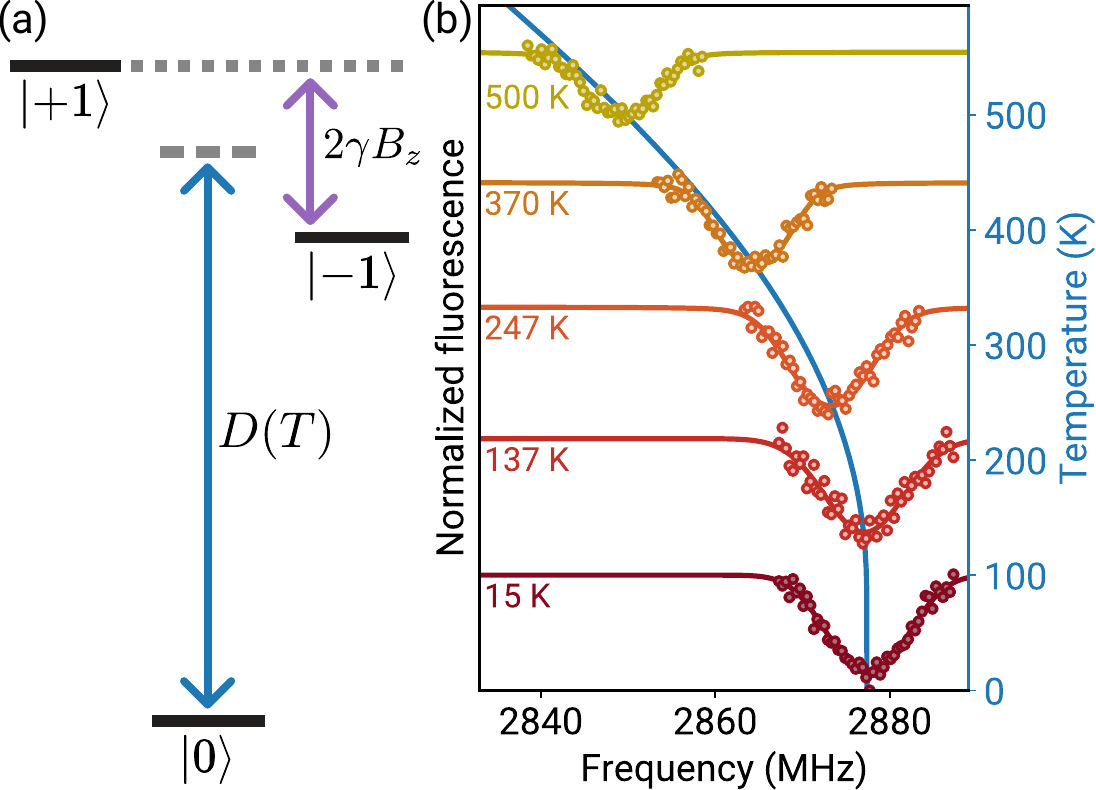}
\caption{\label{fig:intro} 
Temperature-dependent energy levels of the electronic spin of the nitrogen-vacancy center. (a) Electronic ground-state level structure of the negatively charged NV center. In the absence of external fields, the \(\ket{m_s=\pm1}\) spin sub-levels are degenerate and have an energy \(D(T)\) above the \(\ket{m_s=0}\) level, where \(D(T)\) is the the temperature-dependent zero-field splitting. In many applications, the \(\ket{\pm1}\) levels are split by an external magnetic field \(B_{z}\) applied along the NV axis. 
(b) Temperature dependence of an NV center's optically detected magnetic resonance spectrum in the absence of a magnetic field. Data points show the normalized fluorescence from a single NV center under zero applied field after spin polarization through optical pumping and a microwave pulse at the frequency indicated by the \(x\)-axis. Solid lines are fits to double-Voigt profiles \cite{SupplementalMaterial}. Each data set is offset such that the minimum of the fit aligns with the appropriate temperature on the righthand \(y\)-axis. The blue line shows the zero-field splitting as a function of temperature, \(D(T)\), according to a fit of the proposed model given in Eq.~\eqref{d_of_t-simple} to the full set of experimental data. 
}
\end{figure}


\textit{Experimental methods.}---
The NV center electronic ground-state level structure is depicted in Fig.~\ref{fig:intro}a. An external magnetic field \(B_{z}\) is commonly applied along the NV axis to lift the \(\ket{\pm 1}\) degeneracy by \(2 \gamma B_{z}\), where \(\gamma = 2.8\) MHz/G is the NV gyromagnetic ratio. 
However, measurements in this work are conducted using a pulsed optically-detected magnetic resonance (ODMR) sequence under no externally applied field (Fig.~\ref{fig:intro}b). 
The ZFS is recorded as the center frequency of the fit to each spectrum \cite{SupplementalMaterial}.
The experimental apparatus is similar to that described in Ref.~\cite{cambria2022temperature}. The apparatus features dual low- and high-temperature operation modes where the sample temperature is monitored by a resistance temperature detector (RTD) mounted immediately adjacent to the diamond sample. Experiments are conducted using two sets of NV centers within the same high-purity bulk diamond sample ([NV] \(\sim 10^{-3}\) ppb). Low-temperature measurements were conducted with 5 single NV centers comprising set A at temperatures between 15 and 295 K. Unfortunately, after switching from low-temperature to high-temperature operation we were unable to relocate the region with the NV centers in set A. High-temperature measurements were therefore conducted with 5 single NV centers in a different region comprising set B at temperatures between 296 and 500 K. NVs belonging to the same set are within approximately 100 microns of each other, and are over 10 microns below the diamond surface. The two sets are mutually exclusive and from different regions of the diamond. Further experimental details are available in the Supplemental Material \cite{SupplementalMaterial}. 

\begin{figure}[b]
\includegraphics[width=0.48\textwidth]{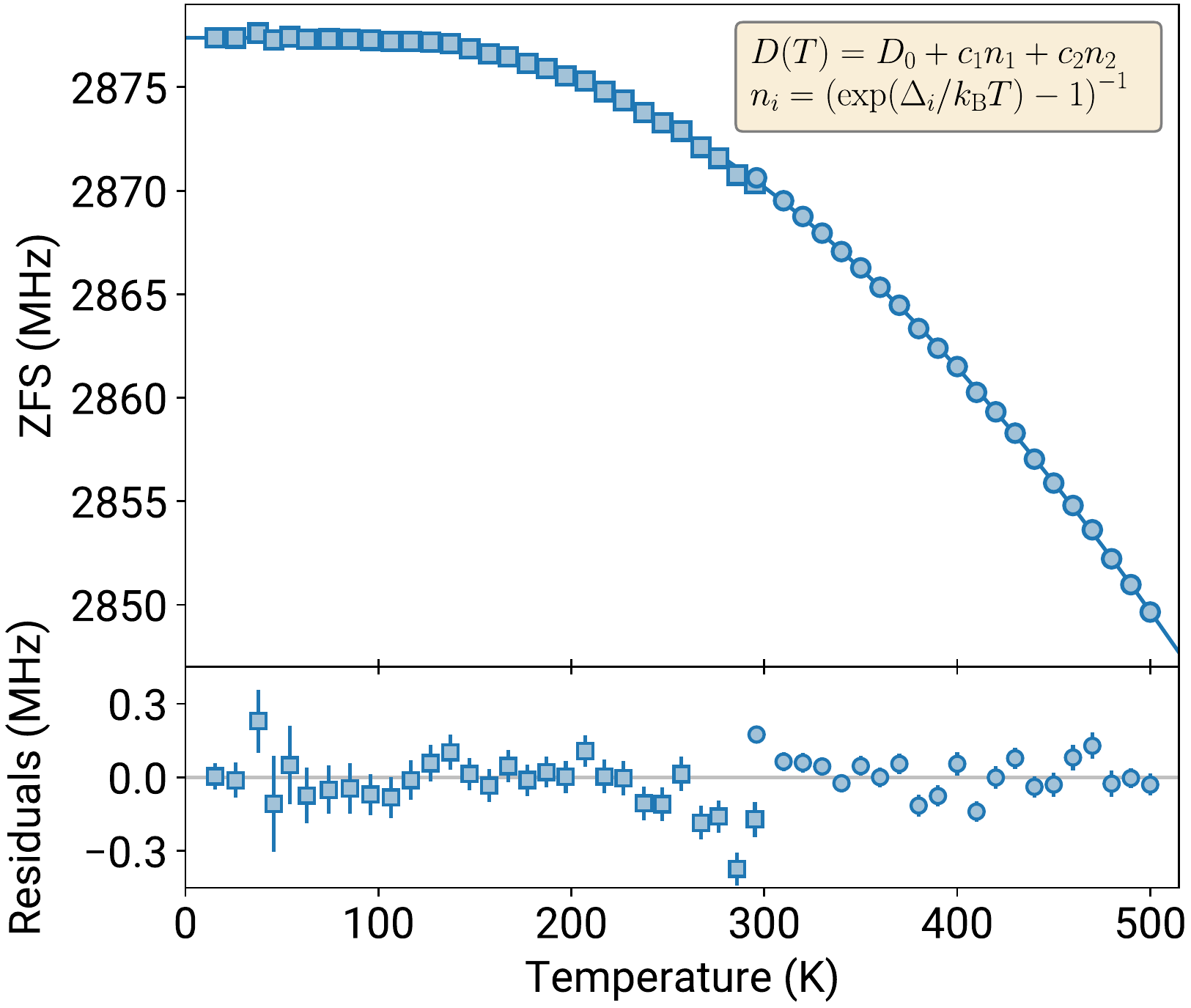}
\caption{\label{fig:zfs_vs_t}
Zero-field splitting as a function of temperature. Top: Data points show the average zero-field ODMR center frequency of several individual NV centers at each temperature. Square (circle) data points indicate measurements conducted in low(high)-temperature operation mode using the same 5 NV centers belonging to set A (B). The two sets of NV centers are mutually exclusive.
Standard error bars are smaller than the data points. Solid line shows fit of the proposed model (Eq.~\eqref{d_of_t-simple}, reproduced in the top right corner of the plot) to the experimental data, with fit parameters presented in Table~\ref{tab:d_of_t-simple} Bottom: Fit residuals. }
\end{figure}


\textit{Results.}--- Figure~\ref{fig:zfs_vs_t} shows the average ZFS recorded for the NV centers belonging to each set (set A - blue filled squares, set B - blue filled circles)  as a function of temperature. 
For reasons that we detail below, the following model provides an excellent fit to the experimental data:
\begin{align}
    D(T) = D_{0} + \sum_{i=1}^M c_{i}n_{i}, \label{d_of_t-all_modes}
\end{align}
where 
\begin{align}
    n_{i} = \frac{1}{e^{\Delta_{i} / k_{\mathrm{B}}T} - 1}.\label{bose}
\end{align}
is the mean occupation number for a phonon mode with energy \(\Delta_{i}\), and $M$ is the total number of discrete phonon modes included in the model. Here \(D(T)\) is the ZFS at a given temperature \(T\), \(D_{0}\) is the ZFS at 0 K, the \(c_{i}\) are coefficients describing the strength of the contribution of each phonon mode to the ZFS, and \(k_{\mathrm{B}}\) is the Boltzmann constant. We find that including only $M=2$ representative phonon modes is sufficient to fully capture the behavior of the ZFS over the entire temperature range measured in this and prior works \cite{SupplementalMaterial}, resulting in the simple analytical expression:
\begin{align}
    D(T) = D_{0} + c_{1}n_{1} + c_{2}n_{2}. \label{d_of_t-simple}
\end{align}

We fit Eq.~\ref{d_of_t-simple} to the data with the quantities \(D_{0}\), \(c_{1,2}\), and \(\Delta_{1,2}\) as free parameters. The values extracted from the fit are presented in Table~\ref{tab:d_of_t-simple}. The energies of the two phonon modes as determined by the fit are $\Delta_1=58.73 \pm 2.32$~meV and $\Delta_2=145.5 \pm 8.4$~meV. These energies are roughly consistent with the highest energies of the diamond acoustic and optical phonon branches respectively, and coincide with the energies of the two representative phonon modes found to dominate spin-phonon relaxation \cite{cambria2022temperature} and the NV optical phonon sideband \cite{davies1974vibronic, collins1987nitrogen, kehayias2013infrared, alkauskas2014first} (\cite{SupplementalMaterial}).

\begin{table*}
    \begin{tabularx}{0.8\textwidth}{|>{\centering\arraybackslash}X >{\centering\arraybackslash}X >{\centering\arraybackslash}X >{\centering\arraybackslash}X >{\centering\arraybackslash}X |} 
     \hline
     \(D_{0}\) (MHz) & \(c_{1}\) (MHz) & \(\Delta_{1}\) (meV) & \(c_{2}\) (MHz) & \(\Delta_{2}\) (meV) \\
     \hline
     2877.38 \(\pm\) 0.03 & -54.91 \(\pm\) 7.35 & 58.73 \(\pm\) 2.32 & -249.6 \(\pm\) 19.4 & 145.5 \(\pm\) 8.4 \\
     \hline
    \end{tabularx}
    \caption{Fit parameters for the ZFS temperature dependence according to the model described by Eq. \eqref{d_of_t-simple}. Values are reported to at least 4 significant figures. Uncertainties are \(1\sigma\).}
    \label{tab:d_of_t-simple}
\end{table*}

The fit residuals are consistent with statistical uncertainties over most of the measured temperature range (Fig.~\ref{fig:zfs_vs_t} bottom). Larger residuals arise near room temperature where the measured ZFS values from sets A and B were found to be offset from each other by approximately 300 kHz. 
The low-temperature measurements made in this work are similarly approximately 300 kHz below those reported by Chen et al.~and Li et al.~in Refs.~\cite{chen2011temperature, li2017temperature}. The offset with respect to these prior measurements is consistent both near room temperature and at very low temperatures where the ZFS has a weak temperature dependence, suggesting that the low ZFS values of the set A NVs may be the result of a small but real physical effect in the diamond resulting in a shift, rather than arising from a systematic error in measurement. 
The ZFS values of the NV centers in set A display more variation than the ZFS values of the NV centers from other regions of the same diamond sample recorded under ambient conditions, suggesting that the local environment of the set A centers may feature large strain or a higher density of charged defects. To investigate this we measured the room temperature ODMR spectra of a number of NV centers in different spatial locations throughout the diamond sample, and identified another region (distinct from the regions containing the set A and set B NV centers) in which most NV centers exhibit MHz-scale splittings of the ZFS line. Taken together, these observations suggest that an individual NV center requires careful calibration in order to function as a temperature sensor with high absolute accuracy. Further discussion and a figure showing the result of the room temperature ODMR survey are available in the Supplemental Material \cite{SupplementalMaterial}. 

\begin{figure}[t]
\includegraphics[width=0.48\textwidth]{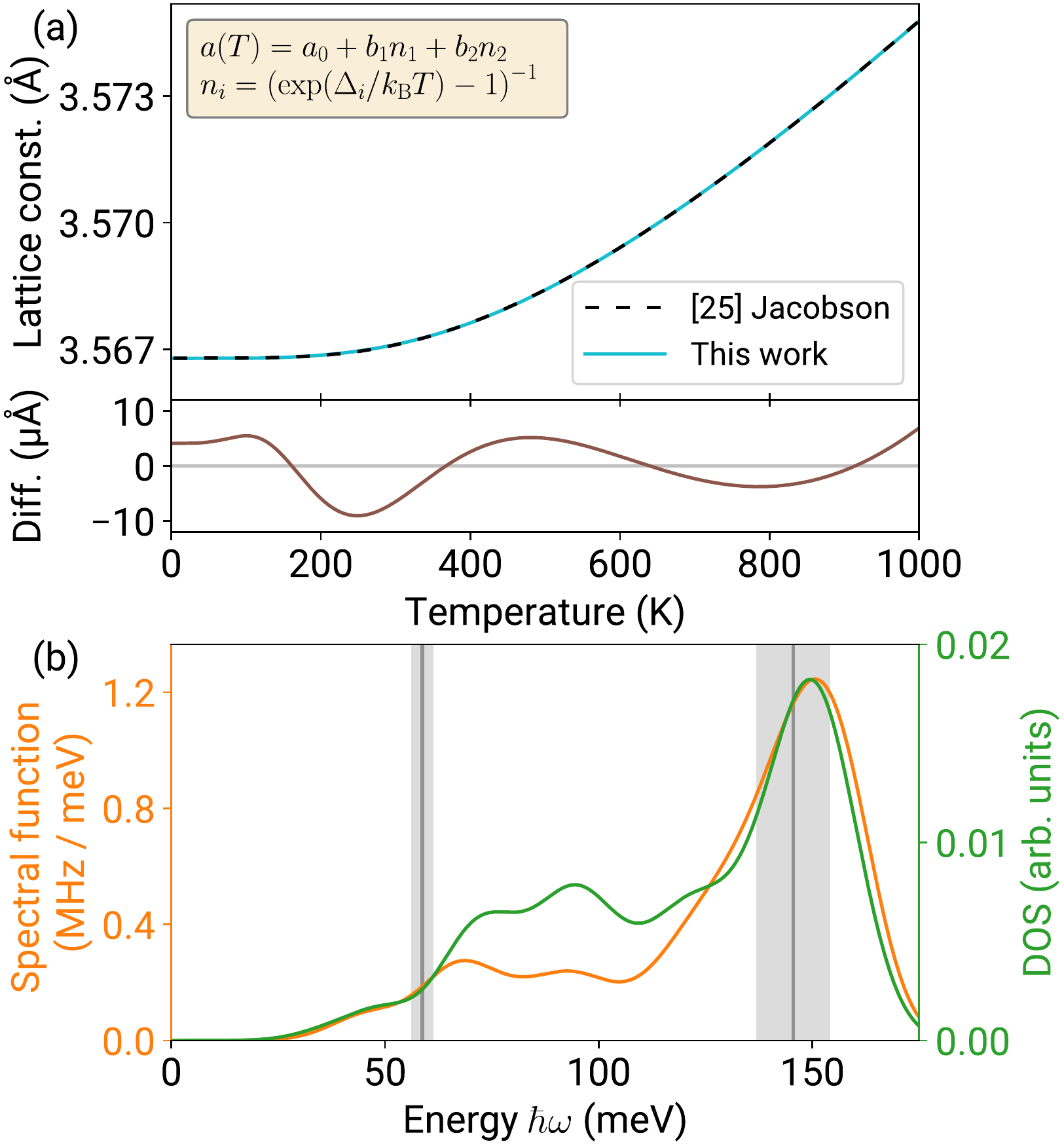}
\caption{\label{fig:model} 
Physical motivation for proposed model of the zero-field splitting temperature dependence. (a) Diamond lattice constant as a function of temperature. Dashed black line indicates the lattice constant of diamond calculated from the results presented in Ref.~\cite{jacobson2019thermal}. Fitting Eq.~\eqref{lattice_constant-simple} (reproduced in the upper left corner of the plot) to this curve yields an excellent approximation of the lattice constant with just three free parameters (cyan line). Importantly, the form of Eq.~\eqref{lattice_constant-simple} matches that of Eq.~\eqref{d_of_t-simple}. Bottom: Fit residuals.
(b) Second-order spin-phonon spectral density (orange) and phonon density of states (DOS, green) calculated from the results of Ref.~\cite{cambria2022temperature}. The spectral density describes the shift of the ZFS induced by phonons at a given energy. The spectral density closely matches the DOS, suggesting that both first-order (thermal expansion) and second-order (``dynamical'') contributions to NV energy shifts can be modeled by a weighted sum of occupation numbers of phonon modes with representative energies. The energies extracted from the fit of Eq.~\ref{d_of_t-simple} to the experimental ZFS data (gray lines and \(\pm 1 \sigma\) intervals) roughly match the locations of prominent features in the spectral density and DOS. 
}
\end{figure}


\textit{Discussion.}---
We now turn to a discussion of the physical origins of Eq.~\ref{d_of_t-all_modes}. 
For a generic solid-state spin system, changes in temperature affect the positions of atoms in the solid which in turn modulate the system's electronic wavefunction, leading the ZFS to acquire a temperature dependence. For the NV center, the ZFS
is the result of the spin-spin interaction between the NV's two unpaired electrons \cite{gali2019ab}. As the electron spin density is highly localized to the three carbon atoms adjacent to the vacancy, the ZFS can be accurately modeled by accounting for the temperature-dependent positions of the atoms nearest the defect \cite{gali2019ab}. Recent \textit{ab initio} work has achieved near-quantitative agreement with experiment by considering the time-averaged atomic positions in a 64 atom supercell up to second order \cite{tang2022first}. Here we adopt a similar premise as a starting point. With \(D_{i}\) indicating the contribution to the ZFS at the \(i\)th order in the atomic displacements, 
\begin{align}
    D(T) = D_{0} + D_{1} + D_{2} + ...
\end{align}
As in Eq.~\eqref{d_of_t-all_modes}, \(D_{0}\) is the ZFS at zero temperature. The linear term is given by
\begin{align}
    D_{1} = \sum_{ik} \frac{\partial D}{\partial u_{ik}} \langle u_{ik} \rangle,\label{d_of_t-1}
\end{align}
where \(\langle \cdot \rangle\) indicates the time average and \(u_{ik}\) is the displacement of the \(i\)th atom in the lattice along the \(k\) axis. The quadratic term is given by
\begin{align}
    D_{2} = \frac{1}{2}\sum_{ii'kk'} \frac{\partial^{2} D}{\partial u_{ik} \partial  u_{i'k'}} \langle u_{ik} u_{i'k'} \rangle.\label{d_of_t-2}
\end{align}
As we explain below, both the first-order and second-order contributions can be written as a weighted sum over a small number of mean phonon occupation numbers, ultimately yielding the proposed model given by Eq.~\eqref{d_of_t-all_modes}. 

Beginning with the first-order contributions, the time-averaged first-order atomic displacements are proportional to the deviation of the lattice constant from its zero-temperature limit:
\begin{align}
    \langle u_{ik} \rangle = \beta_{ik}\left(a(T) - a_{0}\right),\label{atomic_displacements}
\end{align}
where \(a(T)\) and \(a_{0}\) denote the lattice constant at temperature \(T\) and 0 K respectively, and the \(\beta_{ik}\) are coefficients. First-order contributions to the ZFS are therefore proportional to the change in the lattice constant as a function of temperature. To obtain an expression for this quantity, we adopt the quasi-harmonic approximation, in which the crystal's phonon modes are assumed to have volume-dependent frequencies but are otherwise treated harmonically. The quasi-harmonic approximation is widely used to account for thermal expansion in crystals, where the coefficient of thermal expansion is \cite{ashcroft2022solid}
\begin{align}
    \alpha(T) = -\frac{\hbar}{3B} \sum_{ik} \frac{\partial \omega_{ik}}{\partial V} \frac{\partial n_{ik}}{\partial T}. \label{thermal_expansion}
\end{align}
Here, the \(i\)th phonon mode with polarization \(k\) has energy \(\hbar\omega_{ik}\) and mean occupation number \(n_{ik}\). The bulk modulus and volume are denoted \(B\) and \(V\) respectively. 
The frequency derivatives \(\partial \omega_{ik} / \partial V\) are closely related to the Gr\"uneisen parameters that appear in the quasi-harmonic model of thermal expansion \cite{blackman1957thermal, reeber1975thermal}. 
We obtain the lattice constant from the coefficient of thermal expansion using
\begin{align}
    a(T) = a_{0} \exp(\int_{T_{0}}^{T} \alpha(T) \, dT).\label{lattice_constant}
\end{align}
Integrating Eq.~\eqref{thermal_expansion} with \(T_{0}=0\) and neglecting the weak temperature dependence of the bulk modulus, the argument of the exponential becomes a weighted sum over mean occupation numbers. 
After expanding the exponential and dropping higher order terms we obtain
\begin{align}
    a(T) \approx a_{0}\left(1 + \sum_{ik}a_{ik} n_{ik}\right),\label{lattice_constant-all_modes}
\end{align}
where \(a_{ik}=-(\hbar/3B)(\partial\omega_{ik}/\partial V)\). This sum can be simplified by considering only a small number of representative modes with modified coefficients that absorb the effects of the other modes. For diamond, we find that a sum over just two modes displays excellent agreement with measurements of the lattice constant as a function of temperature. Reduced to two modes, Eq.~\eqref{lattice_constant-all_modes} becomes
\begin{align}
    a(T) \approx a_{0} + b_{1} n_{1} + b_{2} n_{2}.\label{lattice_constant-simple}
\end{align}
As in Eq.~\eqref{d_of_t-simple} the \(n_{1,2}\) are the mean occupation numbers of the representative modes. Matching the energies extracted from the ZFS temperature dependence fit, we fix the energies of the modes to \(\Delta_{1}= 58.73\) meV and \(\Delta_{2} = 145.5\) meV. We treat the zero-temperature lattice constant \(a_{0}\) and the coefficients \(b_{1,2}\) as fit parameters. 
To demonstrate that this model accurately captures the temperature dependence of the diamond lattice constant, we fit Eq.~\ref{lattice_constant-all_modes} to a set of lattice constant data generated from the results of Ref.~\cite{jacobson2019thermal}, which accurately matches experimental measurements of the diamond lattice constant. (Further details regarding the generated data are available in the Supplemental Material \cite{SupplementalMaterial}.) As shown in Fig.~\ref{fig:model}a, Eq.~\eqref{lattice_constant-simple} provides an excellent description of the diamond lattice constant with only three free parameters. Over the fit range of 0 to 1000 K, the maximal difference between the two curves is just 9 \(\si{\micro\angstrom}\), or 0.11\% of the change in the lattice constant between 0 and 1000 K.  
The first-order contribution to the NV ZFS temperature dependence is therefore well approximated as 
\begin{align}
    D_{1} \approx c'_{1}n_{1} + c'_{2}n_{2} \label{firstorder-simple}
\end{align}
where the \(c'_{1,2}\) are here simply interpreted as unknown weights. 

\begin{figure}[t]
\includegraphics[width=0.48\textwidth]{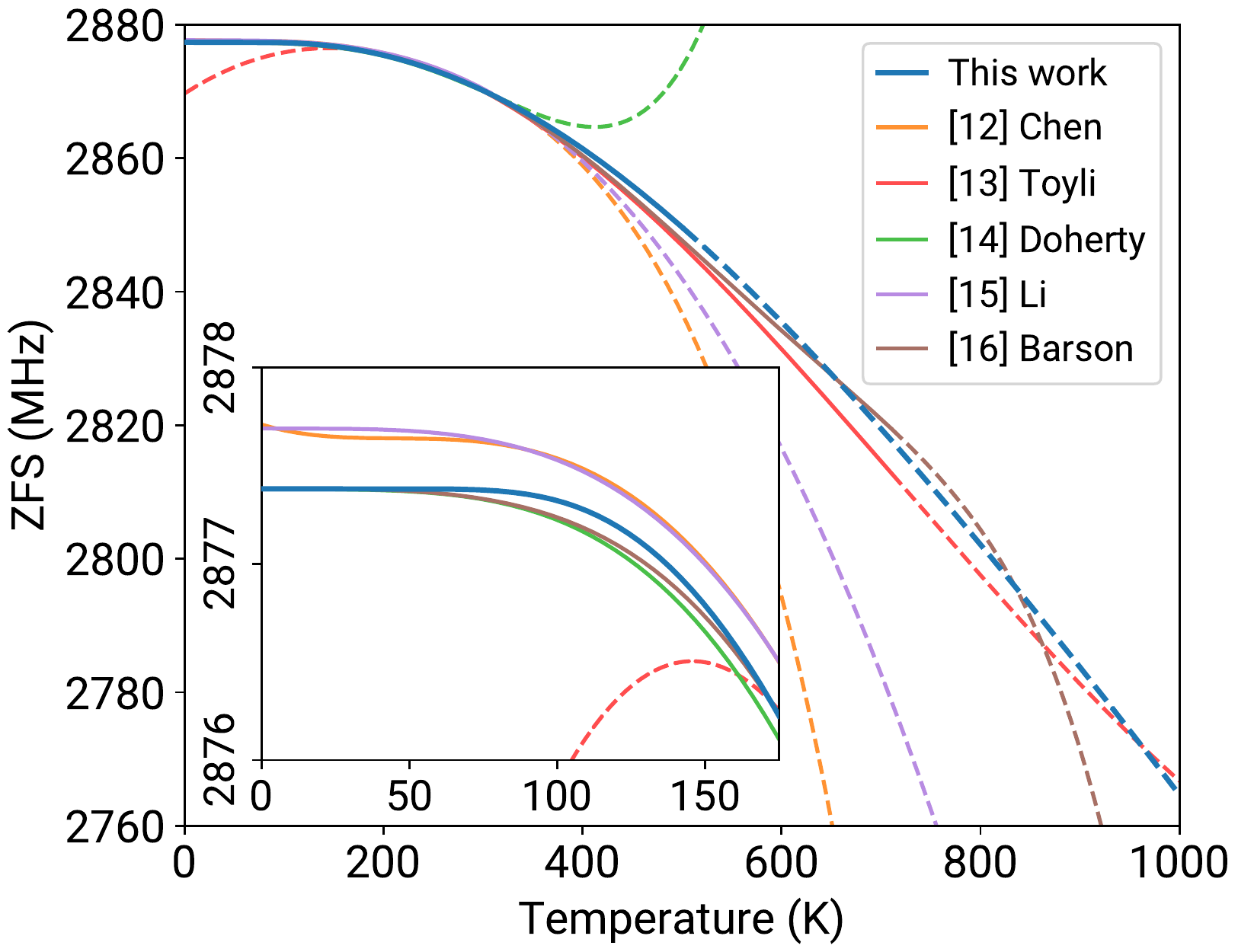}
\caption{\label{fig:comp-models}
Comparison to prior models of the ZFS temperature dependence.
Prior models are reproduced from Refs. \cite{toyli2012measurement, chen2011temperature, doherty2014temperature, li2017temperature, barson2019temperature}. The line for each model turns from solid to dashed outside the temperature range covered by the experimental data on which the model was fit. The model from Ref. \cite{barson2019temperature} is fit to an artificial data set constructed by combining the measurements from Refs. \cite{doherty2014temperature, toyli2012measurement}. 
See Sec.~IV of the Supplemental Material for details regarding prior results \cite{SupplementalMaterial}. Inset: Low-temperature behavior of models.
}
\end{figure}

By writing Eq.~\ref{d_of_t-2} in terms of the normal coordinates \(q_{ik}\) the second-order contribution to the ZFS for a generic system can similarly be expressed as a weighted sum over occupation numbers:
\begin{align}
    D_{2} = \frac{1}{2}\sum_{ii'kk'} \frac{\partial^{2} D}{\partial q_{ik} \partial  q_{i'k'}} \langle q_{ik} q_{i'k'} \rangle.\label{d_of_t-2-sum}
\end{align}
Assuming that the time average of the cross terms is zero, each of the remaining terms \(\langle q_{ik}^{2} \rangle\) is proportional to the mode occupation number \(n_{ik}\) in the quasi-harmonic approximation \cite{tang2022first}. Thus we can write
\begin{align}
    D_{2} = \sum_{ik} \lambda_{ik} n_{ik}
\end{align}
where the \(\lambda_{ik}\) are second-order spin-phonon coupling coefficients proportional to the second derivatives in Eq.~\ref{d_of_t-2-sum}. In the continuum limit, the coupling coefficients are replaced by the second-order spin-phonon spectral density \(S_{2}(\hbar\omega)\) defined such that
\begin{align}
    D_{2} = \int S_{2}(\hbar\omega) \, n(\hbar\omega) \, d(\hbar\omega)
\end{align}
where \(n(\hbar\omega)\) is the mean occupation number at energy \(\hbar\omega\). The NV center spectral density and phonon density of states according to the supercell calculations described in Ref.~\cite{cambria2022temperature} are reproduced in Fig.~\ref{fig:model}a as the orange and green lines respectively. In Ref.~\cite{cambria2022temperature}, it was shown that  NV center spin-phonon relaxation can be reproduced by replacing this spectral density with two representative phonon modes, and we take the same approach here. The similarity between the spectral density and the density of states indicates that this may be done most effectively by retaining the modes associated with a large density of states. Because such modes are also good representative modes for the diamond lattice constant, we expect that the second-order contributions to the ZFS can be expressed by considering the same two representative modes used to describe the first-order contribution as given in Eq.~\ref{firstorder-simple}. Therefore we have
\begin{align}
    D_{2} \approx c''_{1} n_{1} + c''_{2} n_{2}
\end{align}
where the coefficients \(c''_{1,2}\) are again unknown weights. 
Finally combining the contributions up to second order yields
\begin{align}
    D(T) \approx D_{0} + (c'_{1} + c''_{1}) n_{1} + (c'_{2} + c''_{2}) n_{2}
\end{align}
and substituting \(c_{1,2} = c'_{1,2} + c''_{1,2}\), we arrive at Eq.~\eqref{d_of_t-simple}. 

In Fig.~\ref{fig:comp-models} we compare our model to ZFS temperature dependence expressions from prior works that were fit to experimental data spanning a temperature range of at least 100 K. For clarity, the experimental data from the prior works is not shown in the main text plot, but is available in the Supplemental Material \cite{SupplementalMaterial}. The NV center has been coherently manipulated at temperatures up to approximately 1000 K \cite{liu2019coherent}, (at temperatures above 600 K the temperature was verified by recording the ZFS and inverting the ZFS temperature dependence expression from Ref.~\cite{toyli2012measurement},) so we treat 0-1000 K as the full experimentally relevant temperature range. Each model provides a good description of the data on which it was fit (solid regions of lines in Fig.~\ref{fig:comp-models}, see \cite{SupplementalMaterial}), but all models aside from the model proposed in this work diverge at either high or low temperatures outside of the regions they were fit to, which is characteristic of power series. We provide more detailed direct comparisons between the models in the Supplemental Material \cite{SupplementalMaterial}.

The available experimental data sets somewhat disagree with each other in several limits. 
As discussed above, the low-temperature ZFS values measured for this work are approximately 300 kHz lower than those presented in Refs.~\cite{chen2011temperature} and \cite{li2017temperature}. 
The results of Refs.~\cite{toyli2012measurement} and \cite{lourette2022temperature} and our own measurements are in good agreement under ambient conditions, but exhibit slightly different temperature scalings such that the three data sets disagree with each other by several hundred kHz at 400 K \cite{SupplementalMaterial}.
This is most likely due to systematic temperature differentials between the diamond and the temperature sensor in two or potentially all three experiments. A detailed description of how temperature was determined in this work is provided in the Supplemental Material \cite{SupplementalMaterial}. Despite these differences, we emphasize that the ZFS temperature dependence model presented in this work and given by Eq.~\ref{d_of_t-simple} can fully describe all of the available experimental data sets even if the energies of the representative modes are held fixed to the values we report in Table~\ref{tab:d_of_t-simple}, 58.73 and 145.5 meV  \cite{SupplementalMaterial}. 

While our measurements were performed on single NV centers in bulk diamond, we expect that the model of the ZFS temperature dependence presented here should be applicable to high density ensembles of NV centers, near surface NV centers, and NV centers in nanodiamonds, as the characteristic wavelengths of the representative phonon modes are both below 1 nm. The arguments used to arrive at Eq.~\ref{d_of_t-simple} could also be adapted to other phenomena that arise due to spin-lattice interactions, suggesting that similar expressions to Eq.~\eqref{d_of_t-simple} could be replacements for the power series expansions commonly employed in experimental works. The advantages of using a physically motivated expression like Eq.~\ref{d_of_t-simple} are apparent in Fig.~\ref{fig:comp-models}, in which several of the prior models quickly diverge outside the range of experimental on which the models were fit \cite{toyli2012measurement, chen2011temperature}. Other efforts with more sophisticated models still display asymptotic power law behavior that results in similar divergences \cite{li2017temperature, doherty2014temperature, barson2019temperature}. It is likely that the temperature dependence of other important quantities of the NV center, such as the optical zero-phonon line wavelength \cite{chen2011temperature, doherty2014temperature}, hyperfine coupling strengths \cite{xu2022temperature, lourette2022temperature}, and the excited state zero-field splitting \cite{fuchs2008excited, neumann2009excited} could be described by expressions similar to Eq.~\eqref{d_of_t-simple}. Similar models could also be developed for other defect systems, such as divacancy centers in silicon carbide \cite{lin2021temperature} or vacancies in 2D materials \cite{liu2021temperature}. 


\textit{Conclusion.}---
In this work we have presented measurements of the ground-state zero-field splitting (ZFS) in single NV centers in high-purity bulk diamond sample from 15 to 500 K. Our novel analytical model for the NV ZFS, which describes the shift as proportional to the occupation numbers of two representative phonon modes, is in excellent agreement with the experimental data. We explained the physical origins of our model, and suggested that it could replace the power series commonly employed in other works. Finally, we suggested that this model may be readily adapted to several other important properties of the NV center and to other solid-state defects. 

\vspace{\baselineskip}

We gratefully acknowledge Hao Tang, Ariel Rebekah Barr, and Guoqing Wang for enlightening conversations, and Peter Maurer for helpful feedback on our manuscript. Experimental work, data analysis, and theoretical efforts conducted at UW--Madison were supported by the U.S.~Department of Energy, Office of Science, Basic Energy Sciences under Award \#DE-SC0020313. Part of this work by V.~L.~was performed under the auspices of the U.S. Department of Energy at Lawrence Livermore National Laboratory under Contract DE-AC52-07NA27344. A.~N.~acknowledges financial support from Fondecyt Iniciaci\'{o}n No 11220266. A.~G.~acknowledges support from the Department of Defense through the National Defense Science and Engineering Graduate Fellowship (NDSEG) program. \'A.~G.~acknowledges the support from the NKFIH in Hungary for the National Excellence Program (Grant No.\ KKP129866), the Quantum Information National Laboratory (Grant No. 2022-2.1.1-NL-2022-00004), and the EU QuantERA II MAESTRO project and from the European Commission for the QuMicro project (Grant No.\ 101046911). G.~T.~was supported by the János Bolyai Research Scholarship of the Hungarian Academy of Sciences. \'A.~G.~and G.~T.~acknowledge the high-performance computational resources provided by KIFÜ (Governmental Agency for IT Development) institute of Hungary. J.~R.~M.~acknowledges support from ANID-Fondecyt 1221512 and ANID-Anillo ACT192023.. 
\bibliography{MainReferences.bib}

\providecommand{\noopsort}[1]{}\providecommand{\singleletter}[1]{#1}%
\begin{thebibliography}{44}%
\makeatletter
\providecommand \@ifxundefined [1]{%
 \@ifx{#1\undefined}
}%
\providecommand \@ifnum [1]{%
 \ifnum #1\expandafter \@firstoftwo
 \else \expandafter \@secondoftwo
 \fi
}%
\providecommand \@ifx [1]{%
 \ifx #1\expandafter \@firstoftwo
 \else \expandafter \@secondoftwo
 \fi
}%
\providecommand \natexlab [1]{#1}%
\providecommand \enquote  [1]{``#1''}%
\providecommand \bibnamefont  [1]{#1}%
\providecommand \bibfnamefont [1]{#1}%
\providecommand \citenamefont [1]{#1}%
\providecommand \href@noop [0]{\@secondoftwo}%
\providecommand \href [0]{\begingroup \@sanitize@url \@href}%
\providecommand \@href[1]{\@@startlink{#1}\@@href}%
\providecommand \@@href[1]{\endgroup#1\@@endlink}%
\providecommand \@sanitize@url [0]{\catcode `\\12\catcode `\$12\catcode
  `\&12\catcode `\#12\catcode `\^12\catcode `\_12\catcode `\%12\relax}%
\providecommand \@@startlink[1]{}%
\providecommand \@@endlink[0]{}%
\providecommand \url  [0]{\begingroup\@sanitize@url \@url }%
\providecommand \@url [1]{\endgroup\@href {#1}{\urlprefix }}%
\providecommand \urlprefix  [0]{URL }%
\providecommand \Eprint [0]{\href }%
\providecommand \doibase [0]{https://doi.org/}%
\providecommand \selectlanguage [0]{\@gobble}%
\providecommand \bibinfo  [0]{\@secondoftwo}%
\providecommand \bibfield  [0]{\@secondoftwo}%
\providecommand \translation [1]{[#1]}%
\providecommand \BibitemOpen [0]{}%
\providecommand \bibitemStop [0]{}%
\providecommand \bibitemNoStop [0]{.\EOS\space}%
\providecommand \EOS [0]{\spacefactor3000\relax}%
\providecommand \BibitemShut  [1]{\csname bibitem#1\endcsname}%
\let\auto@bib@innerbib\@empty
\bibitem [{\citenamefont {Abendroth}\ \emph {et~al.}(2022)\citenamefont
  {Abendroth}, \citenamefont {Herb}, \citenamefont {Janitz}, \citenamefont
  {Zhu}, \citenamefont {V{\"o}lker},\ and\ \citenamefont
  {Degen}}]{abendroth2022single}%
  \BibitemOpen
  \bibfield  {author} {\bibinfo {author} {\bibfnamefont {J.~M.}\ \bibnamefont
  {Abendroth}}, \bibinfo {author} {\bibfnamefont {K.}~\bibnamefont {Herb}},
  \bibinfo {author} {\bibfnamefont {E.}~\bibnamefont {Janitz}}, \bibinfo
  {author} {\bibfnamefont {T.}~\bibnamefont {Zhu}}, \bibinfo {author}
  {\bibfnamefont {L.~A.}\ \bibnamefont {V{\"o}lker}},\ and\ \bibinfo {author}
  {\bibfnamefont {C.~L.}\ \bibnamefont {Degen}},\ }\bibfield  {title} {\bibinfo
  {title} {Single-nitrogen--vacancy \uppercase{NMR} of amine-functionalized
  diamond surfaces},\ }\href@noop {} {\bibfield  {journal} {\bibinfo  {journal}
  {Nano Letters}\ }\textbf {\bibinfo {volume} {22}},\ \bibinfo {pages} {7294}
  (\bibinfo {year} {2022})}\BibitemShut {NoStop}%
\bibitem [{\citenamefont {Qiu}\ \emph {et~al.}(2022)\citenamefont {Qiu},
  \citenamefont {Hamo}, \citenamefont {Vool}, \citenamefont {Zhou},\ and\
  \citenamefont {Yacoby}}]{qiu2022nanoscale}%
  \BibitemOpen
  \bibfield  {author} {\bibinfo {author} {\bibfnamefont {Z.}~\bibnamefont
  {Qiu}}, \bibinfo {author} {\bibfnamefont {A.}~\bibnamefont {Hamo}}, \bibinfo
  {author} {\bibfnamefont {U.}~\bibnamefont {Vool}}, \bibinfo {author}
  {\bibfnamefont {T.~X.}\ \bibnamefont {Zhou}},\ and\ \bibinfo {author}
  {\bibfnamefont {A.}~\bibnamefont {Yacoby}},\ }\bibfield  {title} {\bibinfo
  {title} {Nanoscale electric field imaging with an ambient scanning quantum
  sensor microscope},\ }\href@noop {} {\bibfield  {journal} {\bibinfo
  {journal} {npj Quantum Information}\ }\textbf {\bibinfo {volume} {8}},\
  \bibinfo {pages} {107} (\bibinfo {year} {2022})}\BibitemShut {NoStop}%
\bibitem [{\citenamefont {Chen}\ \emph {et~al.}(2023)\citenamefont {Chen},
  \citenamefont {Li}, \citenamefont {Yang}, \citenamefont {Ekimov},
  \citenamefont {Bradac}, \citenamefont {Ha}, \citenamefont {Toth},
  \citenamefont {Aharonovich},\ and\ \citenamefont {Tran}}]{chen2023real}%
  \BibitemOpen
  \bibfield  {author} {\bibinfo {author} {\bibfnamefont {Y.}~\bibnamefont
  {Chen}}, \bibinfo {author} {\bibfnamefont {C.}~\bibnamefont {Li}}, \bibinfo
  {author} {\bibfnamefont {T.}~\bibnamefont {Yang}}, \bibinfo {author}
  {\bibfnamefont {E.~A.}\ \bibnamefont {Ekimov}}, \bibinfo {author}
  {\bibfnamefont {C.}~\bibnamefont {Bradac}}, \bibinfo {author} {\bibfnamefont
  {S.~T.}\ \bibnamefont {Ha}}, \bibinfo {author} {\bibfnamefont
  {M.}~\bibnamefont {Toth}}, \bibinfo {author} {\bibfnamefont {I.}~\bibnamefont
  {Aharonovich}},\ and\ \bibinfo {author} {\bibfnamefont {T.~T.}\ \bibnamefont
  {Tran}},\ }\bibfield  {title} {\bibinfo {title} {Real-time ratiometric
  optical nanoscale thermometry},\ }\href@noop {} {\bibfield  {journal}
  {\bibinfo  {journal} {ACS nano}\ }\textbf {\bibinfo {volume} {17}},\ \bibinfo
  {pages} {2725} (\bibinfo {year} {2023})}\BibitemShut {NoStop}%
\bibitem [{\citenamefont {Rovny}\ \emph {et~al.}(2022)\citenamefont {Rovny},
  \citenamefont {Yuan}, \citenamefont {Fitzpatrick}, \citenamefont {Abdalla},
  \citenamefont {Futamura}, \citenamefont {Fox}, \citenamefont {Cambria},
  \citenamefont {Kolkowitz},\ and\ \citenamefont
  {de~Leon}}]{rovny2022nanoscale}%
  \BibitemOpen
  \bibfield  {author} {\bibinfo {author} {\bibfnamefont {J.}~\bibnamefont
  {Rovny}}, \bibinfo {author} {\bibfnamefont {Z.}~\bibnamefont {Yuan}},
  \bibinfo {author} {\bibfnamefont {M.}~\bibnamefont {Fitzpatrick}}, \bibinfo
  {author} {\bibfnamefont {A.~I.}\ \bibnamefont {Abdalla}}, \bibinfo {author}
  {\bibfnamefont {L.}~\bibnamefont {Futamura}}, \bibinfo {author}
  {\bibfnamefont {C.}~\bibnamefont {Fox}}, \bibinfo {author} {\bibfnamefont
  {M.~C.}\ \bibnamefont {Cambria}}, \bibinfo {author} {\bibfnamefont
  {S.}~\bibnamefont {Kolkowitz}},\ and\ \bibinfo {author} {\bibfnamefont
  {N.~P.}\ \bibnamefont {de~Leon}},\ }\bibfield  {title} {\bibinfo {title}
  {Nanoscale covariance magnetometry with diamond quantum sensors},\
  }\href@noop {} {\bibfield  {journal} {\bibinfo  {journal} {Science}\ }\textbf
  {\bibinfo {volume} {378}},\ \bibinfo {pages} {1301} (\bibinfo {year}
  {2022})}\BibitemShut {NoStop}%
\bibitem [{\citenamefont {Bradley}\ \emph {et~al.}(2022)\citenamefont
  {Bradley}, \citenamefont {de~Bone}, \citenamefont {M{\"o}ller}, \citenamefont
  {Baier}, \citenamefont {Degen}, \citenamefont {Loenen}, \citenamefont
  {Bartling}, \citenamefont {Markham}, \citenamefont {Twitchen}, \citenamefont
  {Hanson} \emph {et~al.}}]{bradley2022robust}%
  \BibitemOpen
  \bibfield  {author} {\bibinfo {author} {\bibfnamefont {C.}~\bibnamefont
  {Bradley}}, \bibinfo {author} {\bibfnamefont {S.}~\bibnamefont {de~Bone}},
  \bibinfo {author} {\bibfnamefont {P.}~\bibnamefont {M{\"o}ller}}, \bibinfo
  {author} {\bibfnamefont {S.}~\bibnamefont {Baier}}, \bibinfo {author}
  {\bibfnamefont {M.}~\bibnamefont {Degen}}, \bibinfo {author} {\bibfnamefont
  {S.}~\bibnamefont {Loenen}}, \bibinfo {author} {\bibfnamefont
  {H.}~\bibnamefont {Bartling}}, \bibinfo {author} {\bibfnamefont
  {M.}~\bibnamefont {Markham}}, \bibinfo {author} {\bibfnamefont
  {D.}~\bibnamefont {Twitchen}}, \bibinfo {author} {\bibfnamefont
  {R.}~\bibnamefont {Hanson}}, \emph {et~al.},\ }\bibfield  {title} {\bibinfo
  {title} {Robust quantum-network memory based on spin qubits in isotopically
  engineered diamond},\ }\href@noop {} {\bibfield  {journal} {\bibinfo
  {journal} {npj Quantum Information}\ }\textbf {\bibinfo {volume} {8}},\
  \bibinfo {pages} {122} (\bibinfo {year} {2022})}\BibitemShut {NoStop}%
\bibitem [{\citenamefont {Stas}\ \emph {et~al.}(2022)\citenamefont {Stas},
  \citenamefont {Huan}, \citenamefont {Machielse}, \citenamefont {Knall},
  \citenamefont {Suleymanzade}, \citenamefont {Pingault}, \citenamefont
  {Sutula}, \citenamefont {Ding}, \citenamefont {Knaut}, \citenamefont
  {Assumpcao} \emph {et~al.}}]{stas2022robust}%
  \BibitemOpen
  \bibfield  {author} {\bibinfo {author} {\bibfnamefont {P.-J.}\ \bibnamefont
  {Stas}}, \bibinfo {author} {\bibfnamefont {Y.~Q.}\ \bibnamefont {Huan}},
  \bibinfo {author} {\bibfnamefont {B.}~\bibnamefont {Machielse}}, \bibinfo
  {author} {\bibfnamefont {E.~N.}\ \bibnamefont {Knall}}, \bibinfo {author}
  {\bibfnamefont {A.}~\bibnamefont {Suleymanzade}}, \bibinfo {author}
  {\bibfnamefont {B.}~\bibnamefont {Pingault}}, \bibinfo {author}
  {\bibfnamefont {M.}~\bibnamefont {Sutula}}, \bibinfo {author} {\bibfnamefont
  {S.~W.}\ \bibnamefont {Ding}}, \bibinfo {author} {\bibfnamefont {C.~M.}\
  \bibnamefont {Knaut}}, \bibinfo {author} {\bibfnamefont {D.~R.}\ \bibnamefont
  {Assumpcao}}, \emph {et~al.},\ }\bibfield  {title} {\bibinfo {title} {Robust
  multi-qubit quantum network node with integrated error detection},\
  }\href@noop {} {\bibfield  {journal} {\bibinfo  {journal} {Science}\ }\textbf
  {\bibinfo {volume} {378}},\ \bibinfo {pages} {557} (\bibinfo {year}
  {2022})}\BibitemShut {NoStop}%
\bibitem [{\citenamefont {Kucsko}\ \emph {et~al.}(2013)\citenamefont {Kucsko},
  \citenamefont {Maurer}, \citenamefont {Yao}, \citenamefont {Kubo},
  \citenamefont {Noh}, \citenamefont {Lo}, \citenamefont {Park},\ and\
  \citenamefont {Lukin}}]{kucsko2013nanometre}%
  \BibitemOpen
  \bibfield  {author} {\bibinfo {author} {\bibfnamefont {G.}~\bibnamefont
  {Kucsko}}, \bibinfo {author} {\bibfnamefont {P.~C.}\ \bibnamefont {Maurer}},
  \bibinfo {author} {\bibfnamefont {N.~Y.}\ \bibnamefont {Yao}}, \bibinfo
  {author} {\bibfnamefont {M.}~\bibnamefont {Kubo}}, \bibinfo {author}
  {\bibfnamefont {H.~J.}\ \bibnamefont {Noh}}, \bibinfo {author} {\bibfnamefont
  {P.~K.}\ \bibnamefont {Lo}}, \bibinfo {author} {\bibfnamefont
  {H.}~\bibnamefont {Park}},\ and\ \bibinfo {author} {\bibfnamefont {M.~D.}\
  \bibnamefont {Lukin}},\ }\bibfield  {title} {\bibinfo {title}
  {Nanometre-scale thermometry in a living cell},\ }\href@noop {} {\bibfield
  {journal} {\bibinfo  {journal} {Nature}\ }\textbf {\bibinfo {volume} {500}},\
  \bibinfo {pages} {54} (\bibinfo {year} {2013})}\BibitemShut {NoStop}%
\bibitem [{\citenamefont {Fujiwara}\ \emph {et~al.}(2020)\citenamefont
  {Fujiwara}, \citenamefont {Sun}, \citenamefont {Dohms}, \citenamefont
  {Nishimura}, \citenamefont {Suto}, \citenamefont {Takezawa}, \citenamefont
  {Oshimi}, \citenamefont {Zhao}, \citenamefont {Sadzak}, \citenamefont
  {Umehara} \emph {et~al.}}]{fujiwara2020real}%
  \BibitemOpen
  \bibfield  {author} {\bibinfo {author} {\bibfnamefont {M.}~\bibnamefont
  {Fujiwara}}, \bibinfo {author} {\bibfnamefont {S.}~\bibnamefont {Sun}},
  \bibinfo {author} {\bibfnamefont {A.}~\bibnamefont {Dohms}}, \bibinfo
  {author} {\bibfnamefont {Y.}~\bibnamefont {Nishimura}}, \bibinfo {author}
  {\bibfnamefont {K.}~\bibnamefont {Suto}}, \bibinfo {author} {\bibfnamefont
  {Y.}~\bibnamefont {Takezawa}}, \bibinfo {author} {\bibfnamefont
  {K.}~\bibnamefont {Oshimi}}, \bibinfo {author} {\bibfnamefont
  {L.}~\bibnamefont {Zhao}}, \bibinfo {author} {\bibfnamefont {N.}~\bibnamefont
  {Sadzak}}, \bibinfo {author} {\bibfnamefont {Y.}~\bibnamefont {Umehara}},
  \emph {et~al.},\ }\bibfield  {title} {\bibinfo {title} {Real-time nanodiamond
  thermometry probing in vivo thermogenic responses},\ }\href@noop {}
  {\bibfield  {journal} {\bibinfo  {journal} {Science advances}\ }\textbf
  {\bibinfo {volume} {6}},\ \bibinfo {pages} {eaba9636} (\bibinfo {year}
  {2020})}\BibitemShut {NoStop}%
\bibitem [{\citenamefont {Choi}\ \emph {et~al.}(2020)\citenamefont {Choi},
  \citenamefont {Zhou}, \citenamefont {Landig}, \citenamefont {Wu},
  \citenamefont {Yu}, \citenamefont {Von~Stetina}, \citenamefont {Kucsko},
  \citenamefont {Mango}, \citenamefont {Needleman}, \citenamefont {Samuel}
  \emph {et~al.}}]{choi2020probing}%
  \BibitemOpen
  \bibfield  {author} {\bibinfo {author} {\bibfnamefont {J.}~\bibnamefont
  {Choi}}, \bibinfo {author} {\bibfnamefont {H.}~\bibnamefont {Zhou}}, \bibinfo
  {author} {\bibfnamefont {R.}~\bibnamefont {Landig}}, \bibinfo {author}
  {\bibfnamefont {H.-Y.}\ \bibnamefont {Wu}}, \bibinfo {author} {\bibfnamefont
  {X.}~\bibnamefont {Yu}}, \bibinfo {author} {\bibfnamefont {S.~E.}\
  \bibnamefont {Von~Stetina}}, \bibinfo {author} {\bibfnamefont
  {G.}~\bibnamefont {Kucsko}}, \bibinfo {author} {\bibfnamefont {S.~E.}\
  \bibnamefont {Mango}}, \bibinfo {author} {\bibfnamefont {D.~J.}\ \bibnamefont
  {Needleman}}, \bibinfo {author} {\bibfnamefont {A.~D.}\ \bibnamefont
  {Samuel}}, \emph {et~al.},\ }\bibfield  {title} {\bibinfo {title} {Probing
  and manipulating embryogenesis via nanoscale thermometry and temperature
  control},\ }\href@noop {} {\bibfield  {journal} {\bibinfo  {journal}
  {Proceedings of the National Academy of Sciences}\ }\textbf {\bibinfo
  {volume} {117}},\ \bibinfo {pages} {14636} (\bibinfo {year}
  {2020})}\BibitemShut {NoStop}%
\bibitem [{\citenamefont {Laraoui}\ \emph {et~al.}(2015)\citenamefont
  {Laraoui}, \citenamefont {Aycock-Rizzo}, \citenamefont {Gao}, \citenamefont
  {Lu}, \citenamefont {Riedo},\ and\ \citenamefont
  {Meriles}}]{laraoui2015imaging}%
  \BibitemOpen
  \bibfield  {author} {\bibinfo {author} {\bibfnamefont {A.}~\bibnamefont
  {Laraoui}}, \bibinfo {author} {\bibfnamefont {H.}~\bibnamefont
  {Aycock-Rizzo}}, \bibinfo {author} {\bibfnamefont {Y.}~\bibnamefont {Gao}},
  \bibinfo {author} {\bibfnamefont {X.}~\bibnamefont {Lu}}, \bibinfo {author}
  {\bibfnamefont {E.}~\bibnamefont {Riedo}},\ and\ \bibinfo {author}
  {\bibfnamefont {C.~A.}\ \bibnamefont {Meriles}},\ }\bibfield  {title}
  {\bibinfo {title} {Imaging thermal conductivity with nanoscale resolution
  using a scanning spin probe},\ }\href@noop {} {\bibfield  {journal} {\bibinfo
   {journal} {Nature communications}\ }\textbf {\bibinfo {volume} {6}},\
  \bibinfo {pages} {8954} (\bibinfo {year} {2015})}\BibitemShut {NoStop}%
\bibitem [{\citenamefont {Acosta}\ \emph {et~al.}(2010)\citenamefont {Acosta},
  \citenamefont {Bauch}, \citenamefont {Ledbetter}, \citenamefont {Waxman},
  \citenamefont {Bouchard},\ and\ \citenamefont
  {Budker}}]{acosta2010temperature}%
  \BibitemOpen
  \bibfield  {author} {\bibinfo {author} {\bibfnamefont {V.~M.}\ \bibnamefont
  {Acosta}}, \bibinfo {author} {\bibfnamefont {E.}~\bibnamefont {Bauch}},
  \bibinfo {author} {\bibfnamefont {M.~P.}\ \bibnamefont {Ledbetter}}, \bibinfo
  {author} {\bibfnamefont {A.}~\bibnamefont {Waxman}}, \bibinfo {author}
  {\bibfnamefont {L.-S.}\ \bibnamefont {Bouchard}},\ and\ \bibinfo {author}
  {\bibfnamefont {D.}~\bibnamefont {Budker}},\ }\bibfield  {title} {\bibinfo
  {title} {Temperature dependence of the nitrogen-vacancy magnetic resonance in
  diamond},\ }\href@noop {} {\bibfield  {journal} {\bibinfo  {journal}
  {Physical review letters}\ }\textbf {\bibinfo {volume} {104}},\ \bibinfo
  {pages} {070801} (\bibinfo {year} {2010})}\BibitemShut {NoStop}%
\bibitem [{\citenamefont {Chen}\ \emph {et~al.}(2011)\citenamefont {Chen},
  \citenamefont {Dong}, \citenamefont {Sun}, \citenamefont {Zou}, \citenamefont
  {Cui}, \citenamefont {Han},\ and\ \citenamefont {Guo}}]{chen2011temperature}%
  \BibitemOpen
  \bibfield  {author} {\bibinfo {author} {\bibfnamefont {X.-D.}\ \bibnamefont
  {Chen}}, \bibinfo {author} {\bibfnamefont {C.-H.}\ \bibnamefont {Dong}},
  \bibinfo {author} {\bibfnamefont {F.-W.}\ \bibnamefont {Sun}}, \bibinfo
  {author} {\bibfnamefont {C.-L.}\ \bibnamefont {Zou}}, \bibinfo {author}
  {\bibfnamefont {J.-M.}\ \bibnamefont {Cui}}, \bibinfo {author} {\bibfnamefont
  {Z.-F.}\ \bibnamefont {Han}},\ and\ \bibinfo {author} {\bibfnamefont {G.-C.}\
  \bibnamefont {Guo}},\ }\bibfield  {title} {\bibinfo {title} {Temperature
  dependent energy level shifts of nitrogen-vacancy centers in diamond},\
  }\href@noop {} {\bibfield  {journal} {\bibinfo  {journal} {Applied Physics
  Letters}\ }\textbf {\bibinfo {volume} {99}},\ \bibinfo {pages} {161903}
  (\bibinfo {year} {2011})}\BibitemShut {NoStop}%
\bibitem [{\citenamefont {Toyli}\ \emph {et~al.}(2012)\citenamefont {Toyli},
  \citenamefont {Christle}, \citenamefont {Alkauskas}, \citenamefont {Buckley},
  \citenamefont {Van~de Walle},\ and\ \citenamefont
  {Awschalom}}]{toyli2012measurement}%
  \BibitemOpen
  \bibfield  {author} {\bibinfo {author} {\bibfnamefont {D.}~\bibnamefont
  {Toyli}}, \bibinfo {author} {\bibfnamefont {D.}~\bibnamefont {Christle}},
  \bibinfo {author} {\bibfnamefont {A.}~\bibnamefont {Alkauskas}}, \bibinfo
  {author} {\bibfnamefont {B.}~\bibnamefont {Buckley}}, \bibinfo {author}
  {\bibfnamefont {C.}~\bibnamefont {Van~de Walle}},\ and\ \bibinfo {author}
  {\bibfnamefont {D.}~\bibnamefont {Awschalom}},\ }\bibfield  {title} {\bibinfo
  {title} {Measurement and control of single nitrogen-vacancy center spins
  above 600 \uppercase{K}},\ }\href {https://doi.org/10.1103/PhysRevX.2.031001}
  {\bibfield  {journal} {\bibinfo  {journal} {Physical Review X}\ }\textbf
  {\bibinfo {volume} {2}},\ \bibinfo {pages} {031001} (\bibinfo {year}
  {2012})}\BibitemShut {NoStop}%
\bibitem [{\citenamefont {Doherty}\ \emph
  {et~al.}(2014{\natexlab{a}})\citenamefont {Doherty}, \citenamefont {Acosta},
  \citenamefont {Jarmola}, \citenamefont {Barson}, \citenamefont {Manson},
  \citenamefont {Budker},\ and\ \citenamefont
  {Hollenberg}}]{doherty2014temperature}%
  \BibitemOpen
  \bibfield  {author} {\bibinfo {author} {\bibfnamefont {M.~W.}\ \bibnamefont
  {Doherty}}, \bibinfo {author} {\bibfnamefont {V.~M.}\ \bibnamefont {Acosta}},
  \bibinfo {author} {\bibfnamefont {A.}~\bibnamefont {Jarmola}}, \bibinfo
  {author} {\bibfnamefont {M.~S.}\ \bibnamefont {Barson}}, \bibinfo {author}
  {\bibfnamefont {N.~B.}\ \bibnamefont {Manson}}, \bibinfo {author}
  {\bibfnamefont {D.}~\bibnamefont {Budker}},\ and\ \bibinfo {author}
  {\bibfnamefont {L.~C.}\ \bibnamefont {Hollenberg}},\ }\bibfield  {title}
  {\bibinfo {title} {Temperature shifts of the resonances of the
  \uppercase{NV}\(^{-}\) center in diamond},\ }\href@noop {} {\bibfield
  {journal} {\bibinfo  {journal} {Physical Review B}\ }\textbf {\bibinfo
  {volume} {90}},\ \bibinfo {pages} {041201} (\bibinfo {year}
  {2014}{\natexlab{a}})}\BibitemShut {NoStop}%
\bibitem [{\citenamefont {Li}\ \emph {et~al.}(2017)\citenamefont {Li},
  \citenamefont {Gong}, \citenamefont {Chen}, \citenamefont {Li}, \citenamefont
  {Zhao}, \citenamefont {Dong}, \citenamefont {Guo},\ and\ \citenamefont
  {Sun}}]{li2017temperature}%
  \BibitemOpen
  \bibfield  {author} {\bibinfo {author} {\bibfnamefont {C.-C.}\ \bibnamefont
  {Li}}, \bibinfo {author} {\bibfnamefont {M.}~\bibnamefont {Gong}}, \bibinfo
  {author} {\bibfnamefont {X.-D.}\ \bibnamefont {Chen}}, \bibinfo {author}
  {\bibfnamefont {S.}~\bibnamefont {Li}}, \bibinfo {author} {\bibfnamefont
  {B.-W.}\ \bibnamefont {Zhao}}, \bibinfo {author} {\bibfnamefont
  {Y.}~\bibnamefont {Dong}}, \bibinfo {author} {\bibfnamefont {G.-C.}\
  \bibnamefont {Guo}},\ and\ \bibinfo {author} {\bibfnamefont {F.-W.}\
  \bibnamefont {Sun}},\ }\bibfield  {title} {\bibinfo {title} {Temperature
  dependent energy gap shifts of single color center in diamond based on
  modified \uppercase{V}arshni equation},\ }\href@noop {} {\bibfield  {journal}
  {\bibinfo  {journal} {Diamond and Related Materials}\ }\textbf {\bibinfo
  {volume} {74}},\ \bibinfo {pages} {119} (\bibinfo {year} {2017})}\BibitemShut
  {NoStop}%
\bibitem [{\citenamefont {Barson}\ \emph {et~al.}(2019)\citenamefont {Barson},
  \citenamefont {Reddy}, \citenamefont {Yang}, \citenamefont {Manson},
  \citenamefont {Wrachtrup},\ and\ \citenamefont
  {Doherty}}]{barson2019temperature}%
  \BibitemOpen
  \bibfield  {author} {\bibinfo {author} {\bibfnamefont {M.}~\bibnamefont
  {Barson}}, \bibinfo {author} {\bibfnamefont {P.}~\bibnamefont {Reddy}},
  \bibinfo {author} {\bibfnamefont {S.}~\bibnamefont {Yang}}, \bibinfo {author}
  {\bibfnamefont {N.}~\bibnamefont {Manson}}, \bibinfo {author} {\bibfnamefont
  {J.}~\bibnamefont {Wrachtrup}},\ and\ \bibinfo {author} {\bibfnamefont
  {M.~W.}\ \bibnamefont {Doherty}},\ }\bibfield  {title} {\bibinfo {title}
  {Temperature dependence of the \(^{13}\)\uppercase{C} hyperfine structure of
  the negatively charged nitrogen-vacancy center in diamond},\ }\href@noop {}
  {\bibfield  {journal} {\bibinfo  {journal} {Physical Review B}\ }\textbf
  {\bibinfo {volume} {99}},\ \bibinfo {pages} {094101} (\bibinfo {year}
  {2019})}\BibitemShut {NoStop}%
\bibitem [{\citenamefont {Lourette}\ \emph {et~al.}(2022)\citenamefont
  {Lourette}, \citenamefont {Jarmola}, \citenamefont {Acosta}, \citenamefont
  {Birdwell}, \citenamefont {Budker}, \citenamefont {Doherty}, \citenamefont
  {Ivanov},\ and\ \citenamefont {Malinovsky}}]{lourette2022temperature}%
  \BibitemOpen
  \bibfield  {author} {\bibinfo {author} {\bibfnamefont {S.}~\bibnamefont
  {Lourette}}, \bibinfo {author} {\bibfnamefont {A.}~\bibnamefont {Jarmola}},
  \bibinfo {author} {\bibfnamefont {V.~M.}\ \bibnamefont {Acosta}}, \bibinfo
  {author} {\bibfnamefont {A.~G.}\ \bibnamefont {Birdwell}}, \bibinfo {author}
  {\bibfnamefont {D.}~\bibnamefont {Budker}}, \bibinfo {author} {\bibfnamefont
  {M.~W.}\ \bibnamefont {Doherty}}, \bibinfo {author} {\bibfnamefont
  {T.}~\bibnamefont {Ivanov}},\ and\ \bibinfo {author} {\bibfnamefont {V.~S.}\
  \bibnamefont {Malinovsky}},\ }\bibfield  {title} {\bibinfo {title}
  {Temperature sensitivity of \(^{14}\mathrm{NV}\) and \(^{15}\mathrm{NV}\)
  ground state manifolds},\ }\href@noop {} {\bibfield  {journal} {\bibinfo
  {journal} {arXiv preprint arXiv:2212.12169}\ } (\bibinfo {year}
  {2022})}\BibitemShut {NoStop}%
\bibitem [{\citenamefont {Tang}\ \emph {et~al.}(2023)\citenamefont {Tang},
  \citenamefont {Barr}, \citenamefont {Wang}, \citenamefont {Cappellaro},\ and\
  \citenamefont {Li}}]{tang2022first}%
  \BibitemOpen
  \bibfield  {author} {\bibinfo {author} {\bibfnamefont {H.}~\bibnamefont
  {Tang}}, \bibinfo {author} {\bibfnamefont {A.~R.}\ \bibnamefont {Barr}},
  \bibinfo {author} {\bibfnamefont {G.}~\bibnamefont {Wang}}, \bibinfo {author}
  {\bibfnamefont {P.}~\bibnamefont {Cappellaro}},\ and\ \bibinfo {author}
  {\bibfnamefont {J.}~\bibnamefont {Li}},\ }\bibfield  {title} {\bibinfo
  {title} {First-principles calculation of the temperature-dependent transition
  energies in spin defects},\ }\href
  {https://doi.org/10.1021/acs.jpclett.3c00314} {\bibfield  {journal} {\bibinfo
   {journal} {The Journal of Physical Chemistry Letters}\ }\textbf {\bibinfo
  {volume} {14}},\ \bibinfo {pages} {3266} (\bibinfo {year} {2023})},\ \bibinfo
  {note} {pMID: 36977131},\ \Eprint
  {https://arxiv.org/abs/https://doi.org/10.1021/acs.jpclett.3c00314}
  {https://doi.org/10.1021/acs.jpclett.3c00314} \BibitemShut {NoStop}%
\bibitem [{Sup()}]{SupplementalMaterial}%
  \BibitemOpen
  \href@noop {} {}\bibinfo {note} {See Supplemental Material for additional
  experimental and theoretical details, which includes
  Refs.~\cite{wang2022microwave, ouyang2022temperature, mittiga2018imaging,
  sato2002thermal, doherty2014electronic, reeber1996thermal, Rohatgi2022,
  bosak2005phonon, stoupin2010thermal}.}\BibitemShut {Stop}%
\bibitem [{\citenamefont {Cambria}\ \emph {et~al.}(2022)\citenamefont
  {Cambria}, \citenamefont {Norambuena}, \citenamefont {Dinani}, \citenamefont
  {Thiering}, \citenamefont {Gardill}, \citenamefont {Kemeny}, \citenamefont
  {Li}, \citenamefont {Lordi}, \citenamefont {Gali}, \citenamefont {Maze} \emph
  {et~al.}}]{cambria2022temperature}%
  \BibitemOpen
  \bibfield  {author} {\bibinfo {author} {\bibfnamefont {M.}~\bibnamefont
  {Cambria}}, \bibinfo {author} {\bibfnamefont {A.}~\bibnamefont {Norambuena}},
  \bibinfo {author} {\bibfnamefont {H.}~\bibnamefont {Dinani}}, \bibinfo
  {author} {\bibfnamefont {G.}~\bibnamefont {Thiering}}, \bibinfo {author}
  {\bibfnamefont {A.}~\bibnamefont {Gardill}}, \bibinfo {author} {\bibfnamefont
  {I.}~\bibnamefont {Kemeny}}, \bibinfo {author} {\bibfnamefont
  {Y.}~\bibnamefont {Li}}, \bibinfo {author} {\bibfnamefont {V.}~\bibnamefont
  {Lordi}}, \bibinfo {author} {\bibfnamefont {A.}~\bibnamefont {Gali}},
  \bibinfo {author} {\bibfnamefont {J.}~\bibnamefont {Maze}}, \emph {et~al.},\
  }\bibfield  {title} {\bibinfo {title} {Temperature-dependent phonon-induced
  relaxation of the nitrogen-vacancy spin triplet in diamond},\ }\href@noop {}
  {\bibfield  {journal} {\bibinfo  {journal} {arXiv preprint arXiv:2209.14446}\
  } (\bibinfo {year} {2022})}\BibitemShut {NoStop}%
\bibitem [{\citenamefont {Davies}(1974)}]{davies1974vibronic}%
  \BibitemOpen
  \bibfield  {author} {\bibinfo {author} {\bibfnamefont {G.}~\bibnamefont
  {Davies}},\ }\bibfield  {title} {\bibinfo {title} {Vibronic spectra in
  diamond},\ }\href {https://doi.org/10.1088/0022-3719/7/20/019} {\bibfield
  {journal} {\bibinfo  {journal} {Journal of Physics C: Solid State Physics}\
  }\textbf {\bibinfo {volume} {7}},\ \bibinfo {pages} {3797} (\bibinfo {year}
  {1974})}\BibitemShut {NoStop}%
\bibitem [{\citenamefont {Collins}\ \emph {et~al.}(1987)\citenamefont
  {Collins}, \citenamefont {Stanley},\ and\ \citenamefont
  {Woods}}]{collins1987nitrogen}%
  \BibitemOpen
  \bibfield  {author} {\bibinfo {author} {\bibfnamefont {A.}~\bibnamefont
  {Collins}}, \bibinfo {author} {\bibfnamefont {M.}~\bibnamefont {Stanley}},\
  and\ \bibinfo {author} {\bibfnamefont {G.}~\bibnamefont {Woods}},\ }\bibfield
   {title} {\bibinfo {title} {Nitrogen isotope effects in synthetic diamonds},\
  }\href {https://doi.org/10.1088/0022-3727/20/7/022} {\bibfield  {journal}
  {\bibinfo  {journal} {Journal of Physics D: Applied Physics}\ }\textbf
  {\bibinfo {volume} {20}},\ \bibinfo {pages} {969} (\bibinfo {year}
  {1987})}\BibitemShut {NoStop}%
\bibitem [{\citenamefont {Kehayias}\ \emph {et~al.}(2013)\citenamefont
  {Kehayias}, \citenamefont {Doherty}, \citenamefont {English}, \citenamefont
  {Fischer}, \citenamefont {Jarmola}, \citenamefont {Jensen}, \citenamefont
  {Leefer}, \citenamefont {Hemmer}, \citenamefont {Manson},\ and\ \citenamefont
  {Budker}}]{kehayias2013infrared}%
  \BibitemOpen
  \bibfield  {author} {\bibinfo {author} {\bibfnamefont {P.}~\bibnamefont
  {Kehayias}}, \bibinfo {author} {\bibfnamefont {M.}~\bibnamefont {Doherty}},
  \bibinfo {author} {\bibfnamefont {D.}~\bibnamefont {English}}, \bibinfo
  {author} {\bibfnamefont {R.}~\bibnamefont {Fischer}}, \bibinfo {author}
  {\bibfnamefont {A.}~\bibnamefont {Jarmola}}, \bibinfo {author} {\bibfnamefont
  {K.}~\bibnamefont {Jensen}}, \bibinfo {author} {\bibfnamefont
  {N.}~\bibnamefont {Leefer}}, \bibinfo {author} {\bibfnamefont
  {P.}~\bibnamefont {Hemmer}}, \bibinfo {author} {\bibfnamefont
  {N.}~\bibnamefont {Manson}},\ and\ \bibinfo {author} {\bibfnamefont
  {D.}~\bibnamefont {Budker}},\ }\bibfield  {title} {\bibinfo {title} {Infrared
  absorption band and vibronic structure of the nitrogen-vacancy center in
  diamond},\ }\href {https://doi.org/10.1103/PhysRevB.88.165202} {\bibfield
  {journal} {\bibinfo  {journal} {Physical Review B}\ }\textbf {\bibinfo
  {volume} {88}},\ \bibinfo {pages} {165202} (\bibinfo {year}
  {2013})}\BibitemShut {NoStop}%
\bibitem [{\citenamefont {Alkauskas}\ \emph {et~al.}(2014)\citenamefont
  {Alkauskas}, \citenamefont {Buckley}, \citenamefont {Awschalom},\ and\
  \citenamefont {Van~de Walle}}]{alkauskas2014first}%
  \BibitemOpen
  \bibfield  {author} {\bibinfo {author} {\bibfnamefont {A.}~\bibnamefont
  {Alkauskas}}, \bibinfo {author} {\bibfnamefont {B.~B.}\ \bibnamefont
  {Buckley}}, \bibinfo {author} {\bibfnamefont {D.~D.}\ \bibnamefont
  {Awschalom}},\ and\ \bibinfo {author} {\bibfnamefont {C.~G.}\ \bibnamefont
  {Van~de Walle}},\ }\bibfield  {title} {\bibinfo {title} {First-principles
  theory of the luminescence lineshape for the triplet transition in diamond
  \uppercase{NV} centres},\ }\href
  {https://doi.org/10.1088/1367-2630/16/7/073026} {\bibfield  {journal}
  {\bibinfo  {journal} {New Journal of Physics}\ }\textbf {\bibinfo {volume}
  {16}},\ \bibinfo {pages} {073026} (\bibinfo {year} {2014})}\BibitemShut
  {NoStop}%
\bibitem [{\citenamefont {Jacobson}\ and\ \citenamefont
  {Stoupin}(2019)}]{jacobson2019thermal}%
  \BibitemOpen
  \bibfield  {author} {\bibinfo {author} {\bibfnamefont {P.}~\bibnamefont
  {Jacobson}}\ and\ \bibinfo {author} {\bibfnamefont {S.}~\bibnamefont
  {Stoupin}},\ }\bibfield  {title} {\bibinfo {title} {Thermal expansion
  coefficient of diamond in a wide temperature range},\ }\href@noop {}
  {\bibfield  {journal} {\bibinfo  {journal} {Diamond and Related Materials}\
  }\textbf {\bibinfo {volume} {97}},\ \bibinfo {pages} {107469} (\bibinfo
  {year} {2019})}\BibitemShut {NoStop}%
\bibitem [{\citenamefont {Gali}(2019)}]{gali2019ab}%
  \BibitemOpen
  \bibfield  {author} {\bibinfo {author} {\bibfnamefont {{\'A}.}~\bibnamefont
  {Gali}},\ }\bibfield  {title} {\bibinfo {title} {Ab initio theory of the
  nitrogen-vacancy center in diamond},\ }\href@noop {} {\bibfield  {journal}
  {\bibinfo  {journal} {Nanophotonics}\ }\textbf {\bibinfo {volume} {8}},\
  \bibinfo {pages} {1907} (\bibinfo {year} {2019})}\BibitemShut {NoStop}%
\bibitem [{\citenamefont {Ashcroft}\ and\ \citenamefont
  {Mermin}(2022)}]{ashcroft2022solid}%
  \BibitemOpen
  \bibfield  {author} {\bibinfo {author} {\bibfnamefont {N.~W.}\ \bibnamefont
  {Ashcroft}}\ and\ \bibinfo {author} {\bibfnamefont {N.~D.}\ \bibnamefont
  {Mermin}},\ }\href@noop {} {\emph {\bibinfo {title} {Solid state physics}}}\
  (\bibinfo  {publisher} {Cengage Learning},\ \bibinfo {year}
  {2022})\BibitemShut {NoStop}%
\bibitem [{\citenamefont {Blackman}(1957)}]{blackman1957thermal}%
  \BibitemOpen
  \bibfield  {author} {\bibinfo {author} {\bibfnamefont {M.}~\bibnamefont
  {Blackman}},\ }\bibfield  {title} {\bibinfo {title} {On the thermal expansion
  of solids},\ }\href@noop {} {\bibfield  {journal} {\bibinfo  {journal}
  {Proceedings of the Physical Society. Section B}\ }\textbf {\bibinfo {volume}
  {70}},\ \bibinfo {pages} {827} (\bibinfo {year} {1957})}\BibitemShut
  {NoStop}%
\bibitem [{\citenamefont {Reeber}(1975)}]{reeber1975thermal}%
  \BibitemOpen
  \bibfield  {author} {\bibinfo {author} {\bibfnamefont {R.~R.}\ \bibnamefont
  {Reeber}},\ }\bibfield  {title} {\bibinfo {title} {Thermal expansion of some
  group iv elements and zns},\ }\href@noop {} {\bibfield  {journal} {\bibinfo
  {journal} {physica status solidi (a)}\ }\textbf {\bibinfo {volume} {32}},\
  \bibinfo {pages} {321} (\bibinfo {year} {1975})}\BibitemShut {NoStop}%
\bibitem [{\citenamefont {Liu}\ \emph {et~al.}(2019)\citenamefont {Liu},
  \citenamefont {Feng}, \citenamefont {Wang}, \citenamefont {Li},\ and\
  \citenamefont {Liu}}]{liu2019coherent}%
  \BibitemOpen
  \bibfield  {author} {\bibinfo {author} {\bibfnamefont {G.-Q.}\ \bibnamefont
  {Liu}}, \bibinfo {author} {\bibfnamefont {X.}~\bibnamefont {Feng}}, \bibinfo
  {author} {\bibfnamefont {N.}~\bibnamefont {Wang}}, \bibinfo {author}
  {\bibfnamefont {Q.}~\bibnamefont {Li}},\ and\ \bibinfo {author}
  {\bibfnamefont {R.-B.}\ \bibnamefont {Liu}},\ }\bibfield  {title} {\bibinfo
  {title} {Coherent quantum control of nitrogen-vacancy center spins near 1000
  kelvin},\ }\href {https://doi.org/10.1038/s41467-019-09327-2} {\bibfield
  {journal} {\bibinfo  {journal} {Nature Communications}\ }\textbf {\bibinfo
  {volume} {10}},\ \bibinfo {pages} {1344} (\bibinfo {year}
  {2019})}\BibitemShut {NoStop}%
\bibitem [{\citenamefont {Xu}\ \emph {et~al.}(2022)\citenamefont {Xu},
  \citenamefont {Liu}, \citenamefont {Xie}, \citenamefont {Zhao}, \citenamefont
  {Shi}, \citenamefont {Yu}, \citenamefont {Duan}, \citenamefont {Shi},\ and\
  \citenamefont {Du}}]{xu2022temperature}%
  \BibitemOpen
  \bibfield  {author} {\bibinfo {author} {\bibfnamefont {S.}~\bibnamefont
  {Xu}}, \bibinfo {author} {\bibfnamefont {M.}~\bibnamefont {Liu}}, \bibinfo
  {author} {\bibfnamefont {T.}~\bibnamefont {Xie}}, \bibinfo {author}
  {\bibfnamefont {Z.}~\bibnamefont {Zhao}}, \bibinfo {author} {\bibfnamefont
  {Q.}~\bibnamefont {Shi}}, \bibinfo {author} {\bibfnamefont {P.}~\bibnamefont
  {Yu}}, \bibinfo {author} {\bibfnamefont {C.-K.}\ \bibnamefont {Duan}},
  \bibinfo {author} {\bibfnamefont {F.}~\bibnamefont {Shi}},\ and\ \bibinfo
  {author} {\bibfnamefont {J.}~\bibnamefont {Du}},\ }\bibfield  {title}
  {\bibinfo {title} {Temperature-dependent behaviors of single spin defects in
  solids determined with \uppercase{H}z-level precision},\ }\href@noop {}
  {\bibfield  {journal} {\bibinfo  {journal} {arXiv preprint arXiv:2212.02849}\
  } (\bibinfo {year} {2022})}\BibitemShut {NoStop}%
\bibitem [{\citenamefont {Fuchs}\ \emph {et~al.}(2008)\citenamefont {Fuchs},
  \citenamefont {Dobrovitski}, \citenamefont {Hanson}, \citenamefont {Batra},
  \citenamefont {Weis}, \citenamefont {Schenkel},\ and\ \citenamefont
  {Awschalom}}]{fuchs2008excited}%
  \BibitemOpen
  \bibfield  {author} {\bibinfo {author} {\bibfnamefont {G.}~\bibnamefont
  {Fuchs}}, \bibinfo {author} {\bibfnamefont {V.}~\bibnamefont {Dobrovitski}},
  \bibinfo {author} {\bibfnamefont {R.}~\bibnamefont {Hanson}}, \bibinfo
  {author} {\bibfnamefont {A.}~\bibnamefont {Batra}}, \bibinfo {author}
  {\bibfnamefont {C.}~\bibnamefont {Weis}}, \bibinfo {author} {\bibfnamefont
  {T.}~\bibnamefont {Schenkel}},\ and\ \bibinfo {author} {\bibfnamefont
  {D.}~\bibnamefont {Awschalom}},\ }\bibfield  {title} {\bibinfo {title}
  {Excited-state spectroscopy using single spin manipulation in diamond},\
  }\href@noop {} {\bibfield  {journal} {\bibinfo  {journal} {Physical review
  letters}\ }\textbf {\bibinfo {volume} {101}},\ \bibinfo {pages} {117601}
  (\bibinfo {year} {2008})}\BibitemShut {NoStop}%
\bibitem [{\citenamefont {Neumann}\ \emph {et~al.}(2009)\citenamefont
  {Neumann}, \citenamefont {Kolesov}, \citenamefont {Jacques}, \citenamefont
  {Beck}, \citenamefont {Tisler}, \citenamefont {Batalov}, \citenamefont
  {Rogers}, \citenamefont {Manson}, \citenamefont {Balasubramanian},
  \citenamefont {Jelezko} \emph {et~al.}}]{neumann2009excited}%
  \BibitemOpen
  \bibfield  {author} {\bibinfo {author} {\bibfnamefont {P.}~\bibnamefont
  {Neumann}}, \bibinfo {author} {\bibfnamefont {R.}~\bibnamefont {Kolesov}},
  \bibinfo {author} {\bibfnamefont {V.}~\bibnamefont {Jacques}}, \bibinfo
  {author} {\bibfnamefont {J.}~\bibnamefont {Beck}}, \bibinfo {author}
  {\bibfnamefont {J.}~\bibnamefont {Tisler}}, \bibinfo {author} {\bibfnamefont
  {A.}~\bibnamefont {Batalov}}, \bibinfo {author} {\bibfnamefont
  {L.}~\bibnamefont {Rogers}}, \bibinfo {author} {\bibfnamefont
  {N.}~\bibnamefont {Manson}}, \bibinfo {author} {\bibfnamefont
  {G.}~\bibnamefont {Balasubramanian}}, \bibinfo {author} {\bibfnamefont
  {F.}~\bibnamefont {Jelezko}}, \emph {et~al.},\ }\bibfield  {title} {\bibinfo
  {title} {Excited-state spectroscopy of single \uppercase{NV} defects in
  diamond using optically detected magnetic resonance},\ }\href@noop {}
  {\bibfield  {journal} {\bibinfo  {journal} {New Journal of Physics}\ }\textbf
  {\bibinfo {volume} {11}},\ \bibinfo {pages} {013017} (\bibinfo {year}
  {2009})}\BibitemShut {NoStop}%
\bibitem [{\citenamefont {Lin}\ \emph {et~al.}(2021)\citenamefont {Lin},
  \citenamefont {Yan}, \citenamefont {Li}, \citenamefont {Wang}, \citenamefont
  {Hao}, \citenamefont {Zhou}, \citenamefont {Li}, \citenamefont {You},
  \citenamefont {Xu}, \citenamefont {Li} \emph {et~al.}}]{lin2021temperature}%
  \BibitemOpen
  \bibfield  {author} {\bibinfo {author} {\bibfnamefont {W.-X.}\ \bibnamefont
  {Lin}}, \bibinfo {author} {\bibfnamefont {F.-F.}\ \bibnamefont {Yan}},
  \bibinfo {author} {\bibfnamefont {Q.}~\bibnamefont {Li}}, \bibinfo {author}
  {\bibfnamefont {J.-f.}\ \bibnamefont {Wang}}, \bibinfo {author}
  {\bibfnamefont {Z.-H.}\ \bibnamefont {Hao}}, \bibinfo {author} {\bibfnamefont
  {J.-Y.}\ \bibnamefont {Zhou}}, \bibinfo {author} {\bibfnamefont
  {H.}~\bibnamefont {Li}}, \bibinfo {author} {\bibfnamefont {L.-X.}\
  \bibnamefont {You}}, \bibinfo {author} {\bibfnamefont {J.-S.}\ \bibnamefont
  {Xu}}, \bibinfo {author} {\bibfnamefont {C.-F.}\ \bibnamefont {Li}}, \emph
  {et~al.},\ }\bibfield  {title} {\bibinfo {title} {Temperature dependence of
  divacancy spin coherence in implanted silicon carbide},\ }\href@noop {}
  {\bibfield  {journal} {\bibinfo  {journal} {Physical Review B}\ }\textbf
  {\bibinfo {volume} {104}},\ \bibinfo {pages} {125305} (\bibinfo {year}
  {2021})}\BibitemShut {NoStop}%
\bibitem [{\citenamefont {Liu}\ \emph {et~al.}(2021)\citenamefont {Liu},
  \citenamefont {Li}, \citenamefont {Yang}, \citenamefont {Yu}, \citenamefont
  {Meng}, \citenamefont {Wang}, \citenamefont {Li}, \citenamefont {Guo},
  \citenamefont {Yan}, \citenamefont {Li} \emph {et~al.}}]{liu2021temperature}%
  \BibitemOpen
  \bibfield  {author} {\bibinfo {author} {\bibfnamefont {W.}~\bibnamefont
  {Liu}}, \bibinfo {author} {\bibfnamefont {Z.-P.}\ \bibnamefont {Li}},
  \bibinfo {author} {\bibfnamefont {Y.-Z.}\ \bibnamefont {Yang}}, \bibinfo
  {author} {\bibfnamefont {S.}~\bibnamefont {Yu}}, \bibinfo {author}
  {\bibfnamefont {Y.}~\bibnamefont {Meng}}, \bibinfo {author} {\bibfnamefont
  {Z.-A.}\ \bibnamefont {Wang}}, \bibinfo {author} {\bibfnamefont {Z.-C.}\
  \bibnamefont {Li}}, \bibinfo {author} {\bibfnamefont {N.-J.}\ \bibnamefont
  {Guo}}, \bibinfo {author} {\bibfnamefont {F.-F.}\ \bibnamefont {Yan}},
  \bibinfo {author} {\bibfnamefont {Q.}~\bibnamefont {Li}}, \emph {et~al.},\
  }\bibfield  {title} {\bibinfo {title} {Temperature-dependent energy-level
  shifts of spin defects in hexagonal boron nitride},\ }\href@noop {}
  {\bibfield  {journal} {\bibinfo  {journal} {ACS Photonics}\ }\textbf
  {\bibinfo {volume} {8}},\ \bibinfo {pages} {1889} (\bibinfo {year}
  {2021})}\BibitemShut {NoStop}%
\bibitem [{\citenamefont {Wang}\ \emph {et~al.}(2022)\citenamefont {Wang},
  \citenamefont {Zhang}, \citenamefont {Feng},\ and\ \citenamefont
  {Xing}}]{wang2022microwave}%
  \BibitemOpen
  \bibfield  {author} {\bibinfo {author} {\bibfnamefont {Z.}~\bibnamefont
  {Wang}}, \bibinfo {author} {\bibfnamefont {J.}~\bibnamefont {Zhang}},
  \bibinfo {author} {\bibfnamefont {X.}~\bibnamefont {Feng}},\ and\ \bibinfo
  {author} {\bibfnamefont {L.}~\bibnamefont {Xing}},\ }\bibfield  {title}
  {\bibinfo {title} {Microwave heating effect on diamond samples of
  nitrogen-vacancy centers},\ }\href@noop {} {\bibfield  {journal} {\bibinfo
  {journal} {ACS omega}\ }\textbf {\bibinfo {volume} {7}},\ \bibinfo {pages}
  {31538} (\bibinfo {year} {2022})}\BibitemShut {NoStop}%
\bibitem [{\citenamefont {Ouyang}\ \emph {et~al.}(2022)\citenamefont {Ouyang},
  \citenamefont {Wang}, \citenamefont {Xing}, \citenamefont {Feng},
  \citenamefont {Zhang}, \citenamefont {Ren},\ and\ \citenamefont
  {Yang}}]{ouyang2022temperature}%
  \BibitemOpen
  \bibfield  {author} {\bibinfo {author} {\bibfnamefont {K.}~\bibnamefont
  {Ouyang}}, \bibinfo {author} {\bibfnamefont {Z.}~\bibnamefont {Wang}},
  \bibinfo {author} {\bibfnamefont {L.}~\bibnamefont {Xing}}, \bibinfo {author}
  {\bibfnamefont {X.}~\bibnamefont {Feng}}, \bibinfo {author} {\bibfnamefont
  {J.}~\bibnamefont {Zhang}}, \bibinfo {author} {\bibfnamefont
  {C.}~\bibnamefont {Ren}},\ and\ \bibinfo {author} {\bibfnamefont
  {X.}~\bibnamefont {Yang}},\ }\bibfield  {title} {\bibinfo {title}
  {Temperature dependence of the zero-field splitting parameter of
  nitrogen-vacancy centre ensembles in diamond considering microwave and laser
  heating effect},\ }\href@noop {} {\bibfield  {journal} {\bibinfo  {journal}
  {Measurement Science and Technology}\ }\textbf {\bibinfo {volume} {34}},\
  \bibinfo {pages} {015102} (\bibinfo {year} {2022})}\BibitemShut {NoStop}%
\bibitem [{\citenamefont {Mittiga}\ \emph {et~al.}(2018)\citenamefont
  {Mittiga}, \citenamefont {Hsieh}, \citenamefont {Zu}, \citenamefont {Kobrin},
  \citenamefont {Machado}, \citenamefont {Bhattacharyya}, \citenamefont {Rui},
  \citenamefont {Jarmola}, \citenamefont {Choi}, \citenamefont {Budker} \emph
  {et~al.}}]{mittiga2018imaging}%
  \BibitemOpen
  \bibfield  {author} {\bibinfo {author} {\bibfnamefont {T.}~\bibnamefont
  {Mittiga}}, \bibinfo {author} {\bibfnamefont {S.}~\bibnamefont {Hsieh}},
  \bibinfo {author} {\bibfnamefont {C.}~\bibnamefont {Zu}}, \bibinfo {author}
  {\bibfnamefont {B.}~\bibnamefont {Kobrin}}, \bibinfo {author} {\bibfnamefont
  {F.}~\bibnamefont {Machado}}, \bibinfo {author} {\bibfnamefont
  {P.}~\bibnamefont {Bhattacharyya}}, \bibinfo {author} {\bibfnamefont
  {N.}~\bibnamefont {Rui}}, \bibinfo {author} {\bibfnamefont {A.}~\bibnamefont
  {Jarmola}}, \bibinfo {author} {\bibfnamefont {S.}~\bibnamefont {Choi}},
  \bibinfo {author} {\bibfnamefont {D.}~\bibnamefont {Budker}}, \emph
  {et~al.},\ }\bibfield  {title} {\bibinfo {title} {Imaging the local charge
  environment of nitrogen-vacancy centers in diamond},\ }\href@noop {}
  {\bibfield  {journal} {\bibinfo  {journal} {Physical review letters}\
  }\textbf {\bibinfo {volume} {121}},\ \bibinfo {pages} {246402} (\bibinfo
  {year} {2018})}\BibitemShut {NoStop}%
\bibitem [{\citenamefont {Sato}\ \emph {et~al.}(2002)\citenamefont {Sato},
  \citenamefont {Ohashi}, \citenamefont {Sudoh}, \citenamefont {Haruna},\ and\
  \citenamefont {Maeta}}]{sato2002thermal}%
  \BibitemOpen
  \bibfield  {author} {\bibinfo {author} {\bibfnamefont {T.}~\bibnamefont
  {Sato}}, \bibinfo {author} {\bibfnamefont {K.}~\bibnamefont {Ohashi}},
  \bibinfo {author} {\bibfnamefont {T.}~\bibnamefont {Sudoh}}, \bibinfo
  {author} {\bibfnamefont {K.}~\bibnamefont {Haruna}},\ and\ \bibinfo {author}
  {\bibfnamefont {H.}~\bibnamefont {Maeta}},\ }\bibfield  {title} {\bibinfo
  {title} {Thermal expansion of a high purity synthetic diamond single crystal
  at low temperatures},\ }\href@noop {} {\bibfield  {journal} {\bibinfo
  {journal} {Physical Review B}\ }\textbf {\bibinfo {volume} {65}},\ \bibinfo
  {pages} {092102} (\bibinfo {year} {2002})}\BibitemShut {NoStop}%
\bibitem [{\citenamefont {Doherty}\ \emph
  {et~al.}(2014{\natexlab{b}})\citenamefont {Doherty}, \citenamefont
  {Struzhkin}, \citenamefont {Simpson}, \citenamefont {McGuinness},
  \citenamefont {Meng}, \citenamefont {Stacey}, \citenamefont {Karle},
  \citenamefont {Hemley}, \citenamefont {Manson}, \citenamefont {Hollenberg}
  \emph {et~al.}}]{doherty2014electronic}%
  \BibitemOpen
  \bibfield  {author} {\bibinfo {author} {\bibfnamefont {M.~W.}\ \bibnamefont
  {Doherty}}, \bibinfo {author} {\bibfnamefont {V.~V.}\ \bibnamefont
  {Struzhkin}}, \bibinfo {author} {\bibfnamefont {D.~A.}\ \bibnamefont
  {Simpson}}, \bibinfo {author} {\bibfnamefont {L.~P.}\ \bibnamefont
  {McGuinness}}, \bibinfo {author} {\bibfnamefont {Y.}~\bibnamefont {Meng}},
  \bibinfo {author} {\bibfnamefont {A.}~\bibnamefont {Stacey}}, \bibinfo
  {author} {\bibfnamefont {T.~J.}\ \bibnamefont {Karle}}, \bibinfo {author}
  {\bibfnamefont {R.~J.}\ \bibnamefont {Hemley}}, \bibinfo {author}
  {\bibfnamefont {N.~B.}\ \bibnamefont {Manson}}, \bibinfo {author}
  {\bibfnamefont {L.~C.}\ \bibnamefont {Hollenberg}}, \emph {et~al.},\
  }\bibfield  {title} {\bibinfo {title} {Electronic properties and metrology
  applications of the diamond \uppercase{NV}\(^{-}\) center under pressure},\
  }\href@noop {} {\bibfield  {journal} {\bibinfo  {journal} {Physical review
  letters}\ }\textbf {\bibinfo {volume} {112}},\ \bibinfo {pages} {047601}
  (\bibinfo {year} {2014}{\natexlab{b}})}\BibitemShut {NoStop}%
\bibitem [{\citenamefont {Reeber}\ and\ \citenamefont
  {Wang}(1996)}]{reeber1996thermal}%
  \BibitemOpen
  \bibfield  {author} {\bibinfo {author} {\bibfnamefont {R.~R.}\ \bibnamefont
  {Reeber}}\ and\ \bibinfo {author} {\bibfnamefont {K.}~\bibnamefont {Wang}},\
  }\bibfield  {title} {\bibinfo {title} {Thermal expansion, molar volume and
  specific heat of diamond from 0 to 3000 \uppercase{K}},\ }\href@noop {}
  {\bibfield  {journal} {\bibinfo  {journal} {Journal of Electronic Materials}\
  }\textbf {\bibinfo {volume} {25}},\ \bibinfo {pages} {63} (\bibinfo {year}
  {1996})}\BibitemShut {NoStop}%
\bibitem [{\citenamefont {Rohatgi}(2022)}]{Rohatgi2022}%
  \BibitemOpen
  \bibfield  {author} {\bibinfo {author} {\bibfnamefont {A.}~\bibnamefont
  {Rohatgi}},\ }\href {https://automeris.io/WebPlotDigitizer} {\bibinfo {title}
  {Webplotdigitizer: Version 4.6}} (\bibinfo {year} {2022})\BibitemShut
  {NoStop}%
\bibitem [{\citenamefont {Bosak}\ and\ \citenamefont
  {Krisch}(2005)}]{bosak2005phonon}%
  \BibitemOpen
  \bibfield  {author} {\bibinfo {author} {\bibfnamefont {A.}~\bibnamefont
  {Bosak}}\ and\ \bibinfo {author} {\bibfnamefont {M.}~\bibnamefont {Krisch}},\
  }\bibfield  {title} {\bibinfo {title} {Phonon density of states probed by
  inelastic x-ray scattering},\ }\href@noop {} {\bibfield  {journal} {\bibinfo
  {journal} {Physical Review B}\ }\textbf {\bibinfo {volume} {72}},\ \bibinfo
  {pages} {224305} (\bibinfo {year} {2005})}\BibitemShut {NoStop}%
\bibitem [{\citenamefont {Stoupin}\ and\ \citenamefont
  {Shvyd’ko}(2010)}]{stoupin2010thermal}%
  \BibitemOpen
  \bibfield  {author} {\bibinfo {author} {\bibfnamefont {S.}~\bibnamefont
  {Stoupin}}\ and\ \bibinfo {author} {\bibfnamefont {Y.~V.}\ \bibnamefont
  {Shvyd’ko}},\ }\bibfield  {title} {\bibinfo {title} {Thermal expansion of
  diamond at low temperatures},\ }\href@noop {} {\bibfield  {journal} {\bibinfo
   {journal} {Physical review letters}\ }\textbf {\bibinfo {volume} {104}},\
  \bibinfo {pages} {085901} (\bibinfo {year} {2010})}\BibitemShut {NoStop}%
\end{thebibliography}%


\providecommand{\noopsort}[1]{}\providecommand{\singleletter}[1]{#1}%
\begin{thebibliography}{23}%
\makeatletter
\providecommand \@ifxundefined [1]{%
 \@ifx{#1\undefined}
}%
\providecommand \@ifnum [1]{%
 \ifnum #1\expandafter \@firstoftwo
 \else \expandafter \@secondoftwo
 \fi
}%
\providecommand \@ifx [1]{%
 \ifx #1\expandafter \@firstoftwo
 \else \expandafter \@secondoftwo
 \fi
}%
\providecommand \natexlab [1]{#1}%
\providecommand \enquote  [1]{``#1''}%
\providecommand \bibnamefont  [1]{#1}%
\providecommand \bibfnamefont [1]{#1}%
\providecommand \citenamefont [1]{#1}%
\providecommand \href@noop [0]{\@secondoftwo}%
\providecommand \href [0]{\begingroup \@sanitize@url \@href}%
\providecommand \@href[1]{\@@startlink{#1}\@@href}%
\providecommand \@@href[1]{\endgroup#1\@@endlink}%
\providecommand \@sanitize@url [0]{\catcode `\\12\catcode `\$12\catcode
  `\&12\catcode `\#12\catcode `\^12\catcode `\_12\catcode `\%12\relax}%
\providecommand \@@startlink[1]{}%
\providecommand \@@endlink[0]{}%
\providecommand \url  [0]{\begingroup\@sanitize@url \@url }%
\providecommand \@url [1]{\endgroup\@href {#1}{\urlprefix }}%
\providecommand \urlprefix  [0]{URL }%
\providecommand \Eprint [0]{\href }%
\providecommand \doibase [0]{https://doi.org/}%
\providecommand \selectlanguage [0]{\@gobble}%
\providecommand \bibinfo  [0]{\@secondoftwo}%
\providecommand \bibfield  [0]{\@secondoftwo}%
\providecommand \translation [1]{[#1]}%
\providecommand \BibitemOpen [0]{}%
\providecommand \bibitemStop [0]{}%
\providecommand \bibitemNoStop [0]{.\EOS\space}%
\providecommand \EOS [0]{\spacefactor3000\relax}%
\providecommand \BibitemShut  [1]{\csname bibitem#1\endcsname}%
\let\auto@bib@innerbib\@empty
\bibitem [{\citenamefont {Cambria}\ \emph {et~al.}(2022)\citenamefont
  {Cambria}, \citenamefont {Norambuena}, \citenamefont {Dinani}, \citenamefont
  {Thiering}, \citenamefont {Gardill}, \citenamefont {Kemeny}, \citenamefont
  {Li}, \citenamefont {Lordi}, \citenamefont {Gali}, \citenamefont {Maze} \emph
  {et~al.}}]{cambria2022temperature}%
  \BibitemOpen
  \bibfield  {author} {\bibinfo {author} {\bibfnamefont {M.}~\bibnamefont
  {Cambria}}, \bibinfo {author} {\bibfnamefont {A.}~\bibnamefont {Norambuena}},
  \bibinfo {author} {\bibfnamefont {H.}~\bibnamefont {Dinani}}, \bibinfo
  {author} {\bibfnamefont {G.}~\bibnamefont {Thiering}}, \bibinfo {author}
  {\bibfnamefont {A.}~\bibnamefont {Gardill}}, \bibinfo {author} {\bibfnamefont
  {I.}~\bibnamefont {Kemeny}}, \bibinfo {author} {\bibfnamefont
  {Y.}~\bibnamefont {Li}}, \bibinfo {author} {\bibfnamefont {V.}~\bibnamefont
  {Lordi}}, \bibinfo {author} {\bibfnamefont {A.}~\bibnamefont {Gali}},
  \bibinfo {author} {\bibfnamefont {J.}~\bibnamefont {Maze}}, \emph {et~al.},\
  }\bibfield  {title} {\bibinfo {title} {Temperature-dependent phonon-induced
  relaxation of the nitrogen-vacancy spin triplet in diamond},\ }\href@noop {}
  {\bibfield  {journal} {\bibinfo  {journal} {arXiv preprint arXiv:2209.14446}\
  } (\bibinfo {year} {2022})}\BibitemShut {NoStop}%
\bibitem [{\citenamefont {Wang}\ \emph {et~al.}(2022)\citenamefont {Wang},
  \citenamefont {Zhang}, \citenamefont {Feng},\ and\ \citenamefont
  {Xing}}]{wang2022microwave}%
  \BibitemOpen
  \bibfield  {author} {\bibinfo {author} {\bibfnamefont {Z.}~\bibnamefont
  {Wang}}, \bibinfo {author} {\bibfnamefont {J.}~\bibnamefont {Zhang}},
  \bibinfo {author} {\bibfnamefont {X.}~\bibnamefont {Feng}},\ and\ \bibinfo
  {author} {\bibfnamefont {L.}~\bibnamefont {Xing}},\ }\bibfield  {title}
  {\bibinfo {title} {Microwave heating effect on diamond samples of
  nitrogen-vacancy centers},\ }\href@noop {} {\bibfield  {journal} {\bibinfo
  {journal} {ACS omega}\ }\textbf {\bibinfo {volume} {7}},\ \bibinfo {pages}
  {31538} (\bibinfo {year} {2022})}\BibitemShut {NoStop}%
\bibitem [{\citenamefont {Ouyang}\ \emph {et~al.}(2022)\citenamefont {Ouyang},
  \citenamefont {Wang}, \citenamefont {Xing}, \citenamefont {Feng},
  \citenamefont {Zhang}, \citenamefont {Ren},\ and\ \citenamefont
  {Yang}}]{ouyang2022temperature}%
  \BibitemOpen
  \bibfield  {author} {\bibinfo {author} {\bibfnamefont {K.}~\bibnamefont
  {Ouyang}}, \bibinfo {author} {\bibfnamefont {Z.}~\bibnamefont {Wang}},
  \bibinfo {author} {\bibfnamefont {L.}~\bibnamefont {Xing}}, \bibinfo {author}
  {\bibfnamefont {X.}~\bibnamefont {Feng}}, \bibinfo {author} {\bibfnamefont
  {J.}~\bibnamefont {Zhang}}, \bibinfo {author} {\bibfnamefont
  {C.}~\bibnamefont {Ren}},\ and\ \bibinfo {author} {\bibfnamefont
  {X.}~\bibnamefont {Yang}},\ }\bibfield  {title} {\bibinfo {title}
  {Temperature dependence of the zero-field splitting parameter of
  nitrogen-vacancy centre ensembles in diamond considering microwave and laser
  heating effect},\ }\href@noop {} {\bibfield  {journal} {\bibinfo  {journal}
  {Measurement Science and Technology}\ }\textbf {\bibinfo {volume} {34}},\
  \bibinfo {pages} {015102} (\bibinfo {year} {2022})}\BibitemShut {NoStop}%
\bibitem [{\citenamefont {Mittiga}\ \emph {et~al.}(2018)\citenamefont
  {Mittiga}, \citenamefont {Hsieh}, \citenamefont {Zu}, \citenamefont {Kobrin},
  \citenamefont {Machado}, \citenamefont {Bhattacharyya}, \citenamefont {Rui},
  \citenamefont {Jarmola}, \citenamefont {Choi}, \citenamefont {Budker} \emph
  {et~al.}}]{mittiga2018imaging}%
  \BibitemOpen
  \bibfield  {author} {\bibinfo {author} {\bibfnamefont {T.}~\bibnamefont
  {Mittiga}}, \bibinfo {author} {\bibfnamefont {S.}~\bibnamefont {Hsieh}},
  \bibinfo {author} {\bibfnamefont {C.}~\bibnamefont {Zu}}, \bibinfo {author}
  {\bibfnamefont {B.}~\bibnamefont {Kobrin}}, \bibinfo {author} {\bibfnamefont
  {F.}~\bibnamefont {Machado}}, \bibinfo {author} {\bibfnamefont
  {P.}~\bibnamefont {Bhattacharyya}}, \bibinfo {author} {\bibfnamefont
  {N.}~\bibnamefont {Rui}}, \bibinfo {author} {\bibfnamefont {A.}~\bibnamefont
  {Jarmola}}, \bibinfo {author} {\bibfnamefont {S.}~\bibnamefont {Choi}},
  \bibinfo {author} {\bibfnamefont {D.}~\bibnamefont {Budker}}, \emph
  {et~al.},\ }\bibfield  {title} {\bibinfo {title} {Imaging the local charge
  environment of nitrogen-vacancy centers in diamond},\ }\href@noop {}
  {\bibfield  {journal} {\bibinfo  {journal} {Physical review letters}\
  }\textbf {\bibinfo {volume} {121}},\ \bibinfo {pages} {246402} (\bibinfo
  {year} {2018})}\BibitemShut {NoStop}%
\bibitem [{\citenamefont {Chen}\ \emph {et~al.}(2011)\citenamefont {Chen},
  \citenamefont {Dong}, \citenamefont {Sun}, \citenamefont {Zou}, \citenamefont
  {Cui}, \citenamefont {Han},\ and\ \citenamefont {Guo}}]{chen2011temperature}%
  \BibitemOpen
  \bibfield  {author} {\bibinfo {author} {\bibfnamefont {X.-D.}\ \bibnamefont
  {Chen}}, \bibinfo {author} {\bibfnamefont {C.-H.}\ \bibnamefont {Dong}},
  \bibinfo {author} {\bibfnamefont {F.-W.}\ \bibnamefont {Sun}}, \bibinfo
  {author} {\bibfnamefont {C.-L.}\ \bibnamefont {Zou}}, \bibinfo {author}
  {\bibfnamefont {J.-M.}\ \bibnamefont {Cui}}, \bibinfo {author} {\bibfnamefont
  {Z.-F.}\ \bibnamefont {Han}},\ and\ \bibinfo {author} {\bibfnamefont {G.-C.}\
  \bibnamefont {Guo}},\ }\bibfield  {title} {\bibinfo {title} {Temperature
  dependent energy level shifts of nitrogen-vacancy centers in diamond},\
  }\href@noop {} {\bibfield  {journal} {\bibinfo  {journal} {Applied Physics
  Letters}\ }\textbf {\bibinfo {volume} {99}},\ \bibinfo {pages} {161903}
  (\bibinfo {year} {2011})}\BibitemShut {NoStop}%
\bibitem [{\citenamefont {Toyli}\ \emph {et~al.}(2012)\citenamefont {Toyli},
  \citenamefont {Christle}, \citenamefont {Alkauskas}, \citenamefont {Buckley},
  \citenamefont {Van~de Walle},\ and\ \citenamefont
  {Awschalom}}]{toyli2012measurement}%
  \BibitemOpen
  \bibfield  {author} {\bibinfo {author} {\bibfnamefont {D.}~\bibnamefont
  {Toyli}}, \bibinfo {author} {\bibfnamefont {D.}~\bibnamefont {Christle}},
  \bibinfo {author} {\bibfnamefont {A.}~\bibnamefont {Alkauskas}}, \bibinfo
  {author} {\bibfnamefont {B.}~\bibnamefont {Buckley}}, \bibinfo {author}
  {\bibfnamefont {C.}~\bibnamefont {Van~de Walle}},\ and\ \bibinfo {author}
  {\bibfnamefont {D.}~\bibnamefont {Awschalom}},\ }\bibfield  {title} {\bibinfo
  {title} {Measurement and control of single nitrogen-vacancy center spins
  above 600 \uppercase{K}},\ }\href {https://doi.org/10.1103/PhysRevX.2.031001}
  {\bibfield  {journal} {\bibinfo  {journal} {Physical Review X}\ }\textbf
  {\bibinfo {volume} {2}},\ \bibinfo {pages} {031001} (\bibinfo {year}
  {2012})}\BibitemShut {NoStop}%
\bibitem [{\citenamefont {Doherty}\ \emph
  {et~al.}(2014{\natexlab{a}})\citenamefont {Doherty}, \citenamefont {Acosta},
  \citenamefont {Jarmola}, \citenamefont {Barson}, \citenamefont {Manson},
  \citenamefont {Budker},\ and\ \citenamefont
  {Hollenberg}}]{doherty2014temperature}%
  \BibitemOpen
  \bibfield  {author} {\bibinfo {author} {\bibfnamefont {M.~W.}\ \bibnamefont
  {Doherty}}, \bibinfo {author} {\bibfnamefont {V.~M.}\ \bibnamefont {Acosta}},
  \bibinfo {author} {\bibfnamefont {A.}~\bibnamefont {Jarmola}}, \bibinfo
  {author} {\bibfnamefont {M.~S.}\ \bibnamefont {Barson}}, \bibinfo {author}
  {\bibfnamefont {N.~B.}\ \bibnamefont {Manson}}, \bibinfo {author}
  {\bibfnamefont {D.}~\bibnamefont {Budker}},\ and\ \bibinfo {author}
  {\bibfnamefont {L.~C.}\ \bibnamefont {Hollenberg}},\ }\bibfield  {title}
  {\bibinfo {title} {Temperature shifts of the resonances of the
  \uppercase{NV}\(^{-}\) center in diamond},\ }\href@noop {} {\bibfield
  {journal} {\bibinfo  {journal} {Physical Review B}\ }\textbf {\bibinfo
  {volume} {90}},\ \bibinfo {pages} {041201} (\bibinfo {year}
  {2014}{\natexlab{a}})}\BibitemShut {NoStop}%
\bibitem [{\citenamefont {Li}\ \emph {et~al.}(2017)\citenamefont {Li},
  \citenamefont {Gong}, \citenamefont {Chen}, \citenamefont {Li}, \citenamefont
  {Zhao}, \citenamefont {Dong}, \citenamefont {Guo},\ and\ \citenamefont
  {Sun}}]{li2017temperature}%
  \BibitemOpen
  \bibfield  {author} {\bibinfo {author} {\bibfnamefont {C.-C.}\ \bibnamefont
  {Li}}, \bibinfo {author} {\bibfnamefont {M.}~\bibnamefont {Gong}}, \bibinfo
  {author} {\bibfnamefont {X.-D.}\ \bibnamefont {Chen}}, \bibinfo {author}
  {\bibfnamefont {S.}~\bibnamefont {Li}}, \bibinfo {author} {\bibfnamefont
  {B.-W.}\ \bibnamefont {Zhao}}, \bibinfo {author} {\bibfnamefont
  {Y.}~\bibnamefont {Dong}}, \bibinfo {author} {\bibfnamefont {G.-C.}\
  \bibnamefont {Guo}},\ and\ \bibinfo {author} {\bibfnamefont {F.-W.}\
  \bibnamefont {Sun}},\ }\bibfield  {title} {\bibinfo {title} {Temperature
  dependent energy gap shifts of single color center in diamond based on
  modified \uppercase{V}arshni equation},\ }\href@noop {} {\bibfield  {journal}
  {\bibinfo  {journal} {Diamond and Related Materials}\ }\textbf {\bibinfo
  {volume} {74}},\ \bibinfo {pages} {119} (\bibinfo {year} {2017})}\BibitemShut
  {NoStop}%
\bibitem [{\citenamefont {Barson}\ \emph {et~al.}(2019)\citenamefont {Barson},
  \citenamefont {Reddy}, \citenamefont {Yang}, \citenamefont {Manson},
  \citenamefont {Wrachtrup},\ and\ \citenamefont
  {Doherty}}]{barson2019temperature}%
  \BibitemOpen
  \bibfield  {author} {\bibinfo {author} {\bibfnamefont {M.}~\bibnamefont
  {Barson}}, \bibinfo {author} {\bibfnamefont {P.}~\bibnamefont {Reddy}},
  \bibinfo {author} {\bibfnamefont {S.}~\bibnamefont {Yang}}, \bibinfo {author}
  {\bibfnamefont {N.}~\bibnamefont {Manson}}, \bibinfo {author} {\bibfnamefont
  {J.}~\bibnamefont {Wrachtrup}},\ and\ \bibinfo {author} {\bibfnamefont
  {M.~W.}\ \bibnamefont {Doherty}},\ }\bibfield  {title} {\bibinfo {title}
  {Temperature dependence of the \(^{13}\)\uppercase{C} hyperfine structure of
  the negatively charged nitrogen-vacancy center in diamond},\ }\href@noop {}
  {\bibfield  {journal} {\bibinfo  {journal} {Physical Review B}\ }\textbf
  {\bibinfo {volume} {99}},\ \bibinfo {pages} {094101} (\bibinfo {year}
  {2019})}\BibitemShut {NoStop}%
\bibitem [{\citenamefont {Lourette}\ \emph {et~al.}(2022)\citenamefont
  {Lourette}, \citenamefont {Jarmola}, \citenamefont {Acosta}, \citenamefont
  {Birdwell}, \citenamefont {Budker}, \citenamefont {Doherty}, \citenamefont
  {Ivanov},\ and\ \citenamefont {Malinovsky}}]{lourette2022temperature}%
  \BibitemOpen
  \bibfield  {author} {\bibinfo {author} {\bibfnamefont {S.}~\bibnamefont
  {Lourette}}, \bibinfo {author} {\bibfnamefont {A.}~\bibnamefont {Jarmola}},
  \bibinfo {author} {\bibfnamefont {V.~M.}\ \bibnamefont {Acosta}}, \bibinfo
  {author} {\bibfnamefont {A.~G.}\ \bibnamefont {Birdwell}}, \bibinfo {author}
  {\bibfnamefont {D.}~\bibnamefont {Budker}}, \bibinfo {author} {\bibfnamefont
  {M.~W.}\ \bibnamefont {Doherty}}, \bibinfo {author} {\bibfnamefont
  {T.}~\bibnamefont {Ivanov}},\ and\ \bibinfo {author} {\bibfnamefont {V.~S.}\
  \bibnamefont {Malinovsky}},\ }\bibfield  {title} {\bibinfo {title}
  {Temperature sensitivity of \(^{14}\mathrm{NV}\) and \(^{15}\mathrm{NV}\)
  ground state manifolds},\ }\href@noop {} {\bibfield  {journal} {\bibinfo
  {journal} {arXiv preprint arXiv:2212.12169}\ } (\bibinfo {year}
  {2022})}\BibitemShut {NoStop}%
\bibitem [{\citenamefont {Tang}\ \emph {et~al.}(2023)\citenamefont {Tang},
  \citenamefont {Barr}, \citenamefont {Wang}, \citenamefont {Cappellaro},\ and\
  \citenamefont {Li}}]{tang2022first}%
  \BibitemOpen
  \bibfield  {author} {\bibinfo {author} {\bibfnamefont {H.}~\bibnamefont
  {Tang}}, \bibinfo {author} {\bibfnamefont {A.~R.}\ \bibnamefont {Barr}},
  \bibinfo {author} {\bibfnamefont {G.}~\bibnamefont {Wang}}, \bibinfo {author}
  {\bibfnamefont {P.}~\bibnamefont {Cappellaro}},\ and\ \bibinfo {author}
  {\bibfnamefont {J.}~\bibnamefont {Li}},\ }\bibfield  {title} {\bibinfo
  {title} {First-principles calculation of the temperature-dependent transition
  energies in spin defects},\ }\href
  {https://doi.org/10.1021/acs.jpclett.3c00314} {\bibfield  {journal} {\bibinfo
   {journal} {The Journal of Physical Chemistry Letters}\ }\textbf {\bibinfo
  {volume} {14}},\ \bibinfo {pages} {3266} (\bibinfo {year} {2023})},\ \bibinfo
  {note} {pMID: 36977131},\ \Eprint
  {https://arxiv.org/abs/https://doi.org/10.1021/acs.jpclett.3c00314}
  {https://doi.org/10.1021/acs.jpclett.3c00314} \BibitemShut {NoStop}%
\bibitem [{\citenamefont {Sato}\ \emph {et~al.}(2002)\citenamefont {Sato},
  \citenamefont {Ohashi}, \citenamefont {Sudoh}, \citenamefont {Haruna},\ and\
  \citenamefont {Maeta}}]{sato2002thermal}%
  \BibitemOpen
  \bibfield  {author} {\bibinfo {author} {\bibfnamefont {T.}~\bibnamefont
  {Sato}}, \bibinfo {author} {\bibfnamefont {K.}~\bibnamefont {Ohashi}},
  \bibinfo {author} {\bibfnamefont {T.}~\bibnamefont {Sudoh}}, \bibinfo
  {author} {\bibfnamefont {K.}~\bibnamefont {Haruna}},\ and\ \bibinfo {author}
  {\bibfnamefont {H.}~\bibnamefont {Maeta}},\ }\bibfield  {title} {\bibinfo
  {title} {Thermal expansion of a high purity synthetic diamond single crystal
  at low temperatures},\ }\href@noop {} {\bibfield  {journal} {\bibinfo
  {journal} {Physical Review B}\ }\textbf {\bibinfo {volume} {65}},\ \bibinfo
  {pages} {092102} (\bibinfo {year} {2002})}\BibitemShut {NoStop}%
\bibitem [{\citenamefont {Doherty}\ \emph
  {et~al.}(2014{\natexlab{b}})\citenamefont {Doherty}, \citenamefont
  {Struzhkin}, \citenamefont {Simpson}, \citenamefont {McGuinness},
  \citenamefont {Meng}, \citenamefont {Stacey}, \citenamefont {Karle},
  \citenamefont {Hemley}, \citenamefont {Manson}, \citenamefont {Hollenberg}
  \emph {et~al.}}]{doherty2014electronic}%
  \BibitemOpen
  \bibfield  {author} {\bibinfo {author} {\bibfnamefont {M.~W.}\ \bibnamefont
  {Doherty}}, \bibinfo {author} {\bibfnamefont {V.~V.}\ \bibnamefont
  {Struzhkin}}, \bibinfo {author} {\bibfnamefont {D.~A.}\ \bibnamefont
  {Simpson}}, \bibinfo {author} {\bibfnamefont {L.~P.}\ \bibnamefont
  {McGuinness}}, \bibinfo {author} {\bibfnamefont {Y.}~\bibnamefont {Meng}},
  \bibinfo {author} {\bibfnamefont {A.}~\bibnamefont {Stacey}}, \bibinfo
  {author} {\bibfnamefont {T.~J.}\ \bibnamefont {Karle}}, \bibinfo {author}
  {\bibfnamefont {R.~J.}\ \bibnamefont {Hemley}}, \bibinfo {author}
  {\bibfnamefont {N.~B.}\ \bibnamefont {Manson}}, \bibinfo {author}
  {\bibfnamefont {L.~C.}\ \bibnamefont {Hollenberg}}, \emph {et~al.},\
  }\bibfield  {title} {\bibinfo {title} {Electronic properties and metrology
  applications of the diamond \uppercase{NV}\(^{-}\) center under pressure},\
  }\href@noop {} {\bibfield  {journal} {\bibinfo  {journal} {Physical review
  letters}\ }\textbf {\bibinfo {volume} {112}},\ \bibinfo {pages} {047601}
  (\bibinfo {year} {2014}{\natexlab{b}})}\BibitemShut {NoStop}%
\bibitem [{\citenamefont {Reeber}\ and\ \citenamefont
  {Wang}(1996)}]{reeber1996thermal}%
  \BibitemOpen
  \bibfield  {author} {\bibinfo {author} {\bibfnamefont {R.~R.}\ \bibnamefont
  {Reeber}}\ and\ \bibinfo {author} {\bibfnamefont {K.}~\bibnamefont {Wang}},\
  }\bibfield  {title} {\bibinfo {title} {Thermal expansion, molar volume and
  specific heat of diamond from 0 to 3000 \uppercase{K}},\ }\href@noop {}
  {\bibfield  {journal} {\bibinfo  {journal} {Journal of Electronic Materials}\
  }\textbf {\bibinfo {volume} {25}},\ \bibinfo {pages} {63} (\bibinfo {year}
  {1996})}\BibitemShut {NoStop}%
\bibitem [{\citenamefont {Rohatgi}(2022)}]{Rohatgi2022}%
  \BibitemOpen
  \bibfield  {author} {\bibinfo {author} {\bibfnamefont {A.}~\bibnamefont
  {Rohatgi}},\ }\href {https://automeris.io/WebPlotDigitizer} {\bibinfo {title}
  {Webplotdigitizer: Version 4.6}} (\bibinfo {year} {2022})\BibitemShut
  {NoStop}%
\bibitem [{\citenamefont {Davies}(1974)}]{davies1974vibronic}%
  \BibitemOpen
  \bibfield  {author} {\bibinfo {author} {\bibfnamefont {G.}~\bibnamefont
  {Davies}},\ }\bibfield  {title} {\bibinfo {title} {Vibronic spectra in
  diamond},\ }\href {https://doi.org/10.1088/0022-3719/7/20/019} {\bibfield
  {journal} {\bibinfo  {journal} {Journal of Physics C: Solid State Physics}\
  }\textbf {\bibinfo {volume} {7}},\ \bibinfo {pages} {3797} (\bibinfo {year}
  {1974})}\BibitemShut {NoStop}%
\bibitem [{\citenamefont {Collins}\ \emph {et~al.}(1987)\citenamefont
  {Collins}, \citenamefont {Stanley},\ and\ \citenamefont
  {Woods}}]{collins1987nitrogen}%
  \BibitemOpen
  \bibfield  {author} {\bibinfo {author} {\bibfnamefont {A.}~\bibnamefont
  {Collins}}, \bibinfo {author} {\bibfnamefont {M.}~\bibnamefont {Stanley}},\
  and\ \bibinfo {author} {\bibfnamefont {G.}~\bibnamefont {Woods}},\ }\bibfield
   {title} {\bibinfo {title} {Nitrogen isotope effects in synthetic diamonds},\
  }\href@noop {} {\bibfield  {journal} {\bibinfo  {journal} {Journal of Physics
  D: Applied Physics}\ }\textbf {\bibinfo {volume} {20}},\ \bibinfo {pages}
  {969} (\bibinfo {year} {1987})}\BibitemShut {NoStop}%
\bibitem [{\citenamefont {Kehayias}\ \emph {et~al.}(2013)\citenamefont
  {Kehayias}, \citenamefont {Doherty}, \citenamefont {English}, \citenamefont
  {Fischer}, \citenamefont {Jarmola}, \citenamefont {Jensen}, \citenamefont
  {Leefer}, \citenamefont {Hemmer}, \citenamefont {Manson},\ and\ \citenamefont
  {Budker}}]{kehayias2013infrared}%
  \BibitemOpen
  \bibfield  {author} {\bibinfo {author} {\bibfnamefont {P.}~\bibnamefont
  {Kehayias}}, \bibinfo {author} {\bibfnamefont {M.}~\bibnamefont {Doherty}},
  \bibinfo {author} {\bibfnamefont {D.}~\bibnamefont {English}}, \bibinfo
  {author} {\bibfnamefont {R.}~\bibnamefont {Fischer}}, \bibinfo {author}
  {\bibfnamefont {A.}~\bibnamefont {Jarmola}}, \bibinfo {author} {\bibfnamefont
  {K.}~\bibnamefont {Jensen}}, \bibinfo {author} {\bibfnamefont
  {N.}~\bibnamefont {Leefer}}, \bibinfo {author} {\bibfnamefont
  {P.}~\bibnamefont {Hemmer}}, \bibinfo {author} {\bibfnamefont
  {N.}~\bibnamefont {Manson}},\ and\ \bibinfo {author} {\bibfnamefont
  {D.}~\bibnamefont {Budker}},\ }\bibfield  {title} {\bibinfo {title} {Infrared
  absorption band and vibronic structure of the nitrogen-vacancy center in
  diamond},\ }\href@noop {} {\bibfield  {journal} {\bibinfo  {journal}
  {Physical Review B}\ }\textbf {\bibinfo {volume} {88}},\ \bibinfo {pages}
  {165202} (\bibinfo {year} {2013})}\BibitemShut {NoStop}%
\bibitem [{\citenamefont {Alkauskas}\ \emph {et~al.}(2014)\citenamefont
  {Alkauskas}, \citenamefont {Buckley}, \citenamefont {Awschalom},\ and\
  \citenamefont {Van~de Walle}}]{alkauskas2014first}%
  \BibitemOpen
  \bibfield  {author} {\bibinfo {author} {\bibfnamefont {A.}~\bibnamefont
  {Alkauskas}}, \bibinfo {author} {\bibfnamefont {B.~B.}\ \bibnamefont
  {Buckley}}, \bibinfo {author} {\bibfnamefont {D.~D.}\ \bibnamefont
  {Awschalom}},\ and\ \bibinfo {author} {\bibfnamefont {C.~G.}\ \bibnamefont
  {Van~de Walle}},\ }\bibfield  {title} {\bibinfo {title} {First-principles
  theory of the luminescence lineshape for the triplet transition in diamond
  \uppercase{NV} centres},\ }\href
  {https://doi.org/10.1088/1367-2630/16/7/073026} {\bibfield  {journal}
  {\bibinfo  {journal} {New Journal of Physics}\ }\textbf {\bibinfo {volume}
  {16}},\ \bibinfo {pages} {073026} (\bibinfo {year} {2014})}\BibitemShut
  {NoStop}%
\bibitem [{\citenamefont {Bosak}\ and\ \citenamefont
  {Krisch}(2005)}]{bosak2005phonon}%
  \BibitemOpen
  \bibfield  {author} {\bibinfo {author} {\bibfnamefont {A.}~\bibnamefont
  {Bosak}}\ and\ \bibinfo {author} {\bibfnamefont {M.}~\bibnamefont {Krisch}},\
  }\bibfield  {title} {\bibinfo {title} {Phonon density of states probed by
  inelastic x-ray scattering},\ }\href@noop {} {\bibfield  {journal} {\bibinfo
  {journal} {Physical Review B}\ }\textbf {\bibinfo {volume} {72}},\ \bibinfo
  {pages} {224305} (\bibinfo {year} {2005})}\BibitemShut {NoStop}%
\bibitem [{\citenamefont {Jacobson}\ and\ \citenamefont
  {Stoupin}(2019)}]{jacobson2019thermal}%
  \BibitemOpen
  \bibfield  {author} {\bibinfo {author} {\bibfnamefont {P.}~\bibnamefont
  {Jacobson}}\ and\ \bibinfo {author} {\bibfnamefont {S.}~\bibnamefont
  {Stoupin}},\ }\bibfield  {title} {\bibinfo {title} {Thermal expansion
  coefficient of diamond in a wide temperature range},\ }\href@noop {}
  {\bibfield  {journal} {\bibinfo  {journal} {Diamond and Related Materials}\
  }\textbf {\bibinfo {volume} {97}},\ \bibinfo {pages} {107469} (\bibinfo
  {year} {2019})}\BibitemShut {NoStop}%
\bibitem [{\citenamefont {Stoupin}\ and\ \citenamefont
  {Shvyd’ko}(2010)}]{stoupin2010thermal}%
  \BibitemOpen
  \bibfield  {author} {\bibinfo {author} {\bibfnamefont {S.}~\bibnamefont
  {Stoupin}}\ and\ \bibinfo {author} {\bibfnamefont {Y.~V.}\ \bibnamefont
  {Shvyd’ko}},\ }\bibfield  {title} {\bibinfo {title} {Thermal expansion of
  diamond at low temperatures},\ }\href@noop {} {\bibfield  {journal} {\bibinfo
   {journal} {Physical review letters}\ }\textbf {\bibinfo {volume} {104}},\
  \bibinfo {pages} {085901} (\bibinfo {year} {2010})}\BibitemShut {NoStop}%
\bibitem [{\citenamefont {Liu}\ \emph {et~al.}(2019)\citenamefont {Liu},
  \citenamefont {Feng}, \citenamefont {Wang}, \citenamefont {Li},\ and\
  \citenamefont {Liu}}]{liu2019coherent}%
  \BibitemOpen
  \bibfield  {author} {\bibinfo {author} {\bibfnamefont {G.-Q.}\ \bibnamefont
  {Liu}}, \bibinfo {author} {\bibfnamefont {X.}~\bibnamefont {Feng}}, \bibinfo
  {author} {\bibfnamefont {N.}~\bibnamefont {Wang}}, \bibinfo {author}
  {\bibfnamefont {Q.}~\bibnamefont {Li}},\ and\ \bibinfo {author}
  {\bibfnamefont {R.-B.}\ \bibnamefont {Liu}},\ }\bibfield  {title} {\bibinfo
  {title} {Coherent quantum control of nitrogen-vacancy center spins near 1000
  kelvin},\ }\href {https://doi.org/10.1038/s41467-019-09327-2} {\bibfield
  {journal} {\bibinfo  {journal} {Nature Communications}\ }\textbf {\bibinfo
  {volume} {10}},\ \bibinfo {pages} {1344} (\bibinfo {year}
  {2019})}\BibitemShut {NoStop}%
\end{thebibliography}%

\end{document}


\preprint{APS/123-QED}

\title{Supplemental Material for ``A physically motivated analytical expression for the temperature dependence of the zero-field splitting of the nitrogen-vacancy center in diamond''}

\author{M.~C.~Cambria}
 
\affiliation{Department of Physics, University of Wisconsin, Madison, WI 53706, USA}

\author{G.~Thiering}

\affiliation{Wigner Research Centre for Physics, P.O. Box 49, 1525 Budapest, Hungary}

\author{A.~Norambuena}
 
\affiliation{Centro de Optica e Informaci\'on Cu\'antica, Universidad Mayor, Camino la Piramide 5750, Huechuraba, Santiago, Chile}

\author{H.~T.~Dinani}
 
\affiliation{Escuela de Ingenier\'{i}a, Facultad de Ciencias, Ingenier\'{i}a y Tecnolog\'{i}a, Universidad Mayor, Santiago, Chile}

\author{A.~Gardill}

\author{I.~Kemeny}
 
\affiliation{Department of Physics, University of Wisconsin, Madison, WI 53706, USA}

\author{V.~Lordi}
 
\affiliation{Lawrence Livermore National Laboratory, Livermore, CA, 94551, USA}

\author{\'A.~Gali}

\affiliation{Wigner Research Centre for Physics, P.O. Box 49, 1525 Budapest, Hungary}

\affiliation{Department of Atomic Physics, Institute of Physics, Budapest University of Technology and Economics, M\H{u}egyetem rakpart 3., 1111 Budapest, Hungary}

\author{J.~R.~Maze}
 
\affiliation{Instituto de F\'isica, Pontificia Universidad Cat\'olica de Chile, Casilla 306, Santiago, Chile}

\affiliation{Centro de Investigaci\'on en Nanotecnolog\'ia y Materiales Avanzados, Pontificia Universidad Cat\'olica de Chile, Santiago, Chile}

\author{S.~Kolkowitz}
\email{kolkowitz@berkeley.edu}
 
\affiliation{Department of Physics, University of Wisconsin, Madison, WI 53706, USA}

\affiliation{Department of Physics, University of California, Berkeley, CA 94720, USA
}%

\maketitle

\section{Experimental details}

The experiments in this work were conducted using a modified version of the apparatus described previously in Ref.~\cite{cambria2022temperature}. In this section we will describe the main features of the apparatus and how it differs from the one used in Ref.~\cite{cambria2022temperature}. The apparatus consists of a home-built confocal microscope with low- and high-temperature operation modes. In low-temperature operation mode (15-295 K), temperature control is provided by an attocube attoDRY800 closed-cycle cryostat. In order to minimize the gradient between the temperature sensor and the diamond sample, an additional sensor (Lakeshore Cernox CX-1050-SD-HT-1.4L-Q) was installed on top of the cold finger immediately adjacent to the sample. The additional temperature sensor was read out via a 4-wire measurement performed with a Lakeshore 218 monitor. The sensor and diamond sample were adhered to the sample using Ted Pella Leitsilber 200 silver paste. We note that a gradual buildup of ice on the diamond sample below 70 K led to increased optical background and resulted in the comparatively large error bars apparent in measurements near 50 K. In high-temperature operation mode (296-500 K), temperature control is provided by a homebuilt copper hot plate with custom PID control software. The hot plate temperature sensor (TE Connectivity PT100 NB-PTCO-024) was positioned immediately adjacent to the sample and read out via a 4-wire measurement performed with a multimeter (Multicomp Pro MP730028). The temperature sensor and diamond sample were adhered to the hot plate using Ted Pella PELCO High Performance Nickel Paste. In both operation modes, temperature readings were logged once every 2 seconds while an optically-detected magnetic resonance (ODMR) experiment is running. The reported temperature for each data point is the average of the readings over the last 10 minutes of the experiment. The standard deviations of the temperatures recorded during this interval were within 120 mK for all experiments. Peak-to-peak variations were within 500 mK. Because the temperature stability becomes worse as the temperature increases from the base temperature in either operation mode, these values indicate the stability at the highest temperatures reached in either low- or high-temperature operation mode. In both modes, temperatures were roughly \(4\times\) more stable at the lower end of the mode's temperature range versus the higher end. In order to minimize the temperature difference between the sample and the sensor, microwaves were delivered by a wire loop (AWG 36 magnet wire, loop radius \(\appr1.5\) mm) suspended above the diamond sample. 

The pulsed ODMR sequence performed in this work consists of optical polarization (\(\appr 3\) mW at 532 nm for 2 \si{\micro\second}) and a microwave \(\pi\) pulse (variable frequency, \(200\) ns) before optical spin state readout (\(\appr 3\) mW at 532 nm for \(\appr 400\) \si{\nano\second}). The sequence was repeated several hundred thousand times to sufficiently average down the signal. The coordinates of the NV being measured were updated roughly once every two minutes so that the NV remained in focus. The total measurement time varied depending on how many repetitions of the sequence are required, but was between 15 and 25 minutes. 

The diamond sample used in this work was grown via chemical-vapor-deposition by Element Six and displays a native, as-grown NV concentration of approximately \(10^{-3}\) ppb in each of the two regions that were studied. This sample was previously used to characterize spin-lattice relaxation of the NV center ground-state electronic spin in Ref.~\cite{cambria2022temperature}, where it was identified as ``Sample B.''

A potential systematic in temperature dependence studies is measurement-induced heating \cite{wang2022microwave, ouyang2022temperature}, in which the measurement itself places an additional heat load on the sample. If this load is not shared by the temperature sensor, then the resulting temperature difference between the sample and the sensor is a systematic error. We characterized measurement-induced heating by running the ODMR pulse sequence with the laser and microwaves focused directly on the temperature sensor next to the diamond sample in low-temperature operation mode where the heating effect is most significant. At 100 K, the sensor registered a temperature increase of \(\appr 0.6\) K. At 200 K, this was reduced to \(\appr 0.3\) K, and at 286 K the heating was \(\appr 0.1\) K. The rise time of the heating once measurement began was \(\appr 10\) seconds, much shorter than the total measurement duration of at least 15 minutes. While this heating effect is partially accounted for by having the temperature sensor very close to the diamond sample, the effect likely induces a small temperature difference between the sensor and the diamond. We can bound the magnitude of this systematic by assuming that only the diamond sample experiences measurement-induced heating and that the reading of the temperature sensor is the same whether or not a measurement is underway on the adjacent diamond sample. Using the fit model of the ZFS temperature dependence from the main text, we calculate that a +0.6 K temperature change at 100 K induces a -2 kHz ZFS shift, a +0.3 K temperature change at 200 K induces a -11 kHz ZFS shift and a +0.1 K temperature change at 286 K induces a -7 kHz ZFS shift. These values are all nearly an order of magnitude below the typical statistical errors recorded for a given measurement. For completeness we note that heating effects below 100 K induce a negligible ZFS shift according to our model. For example, a +10 K temperature change at 50 K induces a ZFS shift below 1 kHz. 

\begin{figure}[b]
\includegraphics[width=0.98\textwidth]{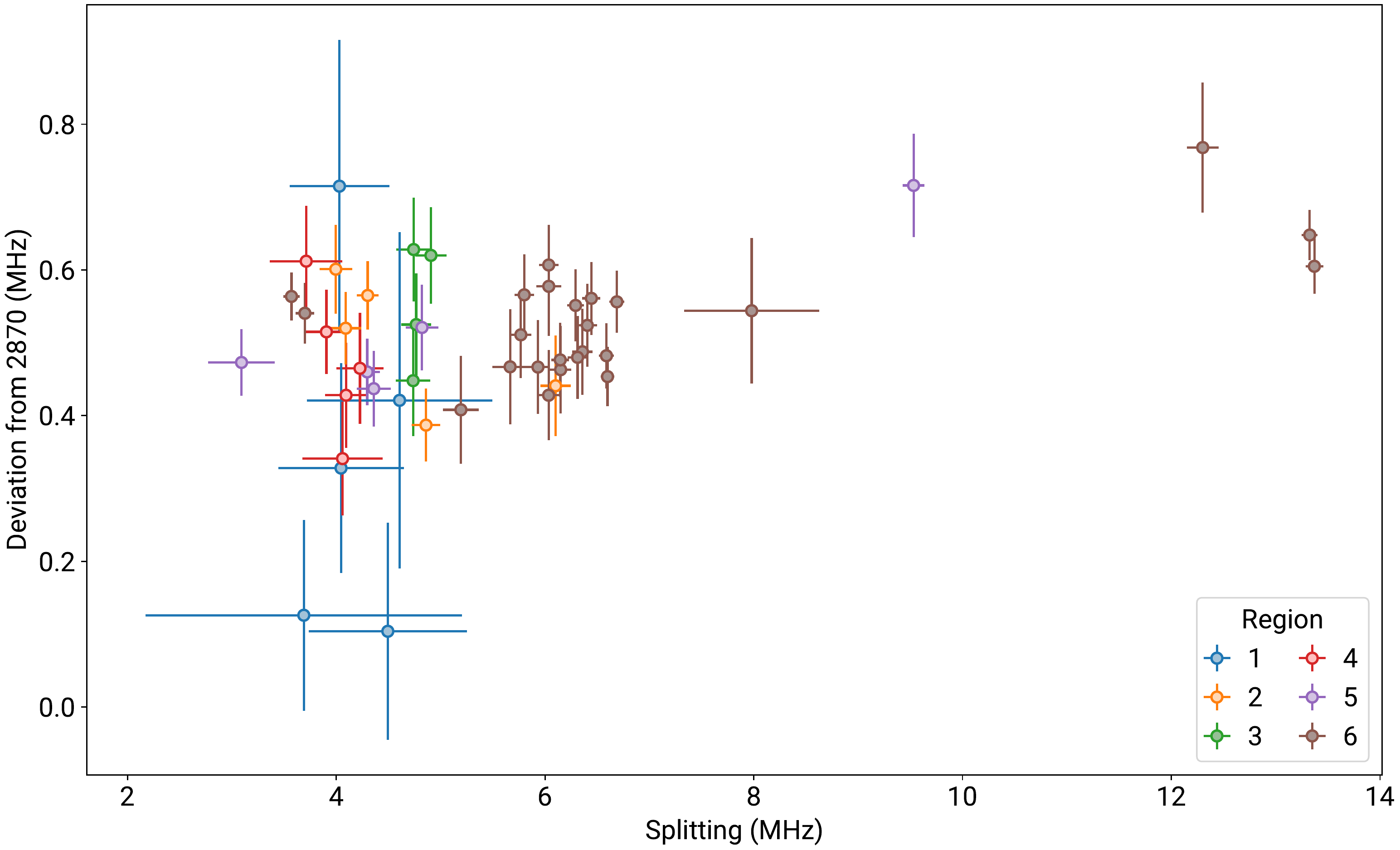}
\caption{\label{fig:zfs_survey}
Near-room temperature zero-field ODMR survey of NV centers from six different regions separated by hundreds of microns in the same diamond sample. All NVs are several tens of microns below the diamond surface. NV centers from region 1 (2) are the set A (B) NVs introduced in the main text. Centers from regions 3-6 were not measured over a broad temperature range. NVs in regions 2-6 were measured under ambient conditions (296 K) in high-temperature operation mode. NVs in region 1 were measured at 295.1 K in low-temperature operation mode. To account for this difference in temperature, the deviations for region 1 NVs are shifted by -62 kHz, which is the difference in ZFS predicted by the model described in the main text between 295.1 and 296 K. After completion of measurements in the cryostat, we attempted to relocate the set A NV centers so that we could re-characterize them under ambient conditions, but were ultimately unable to find them. Note that splittings below the ODMR linewidth (\(\appr 5\) MHz) are possibly artifacts of the ODMR spectrum fit. Visual inspection of the ODMR data sets with splittings below 5 MHz do not display two clearly resolved peaks. At larger splittings, however, two distinct peaks are apparent. 
}
\end{figure}

\section{ODMR spectrum fitting}

Each ODMR spectra is fit with a sum of two identical Voigt profiles that have different center frequencies, which we refer to as a double-Voigt profile. A Voigt profile is the convolution of a Gaussian and a Lorentzian. The parameters that define a single Voigt profile are contrast, center frequency, Gaussian width, and Lorentzian width. If the Gaussian (Lorentzian) width is set to zero, then the Voigt profile is equivalent to a Lorentzian (Gaussian) profile. In principle, because pulsed ODMR is coherent, the ODMR spectrum could be fit by solving for the NV center's Rabi lineshape under zero magnetic field. In practice, this is complicated by the several physical factors that affect the lineshape, including the relevance of all three NV levels at zero field, hyperfine interactions, local electric field and strain, and microwave polarization. We find that the fully empirical and comparatively simple double-Voigt profile provides an excellent description of all the data collected for this work. The profile is parameterized in terms of the ZFS \(D\) and a splitting \(S\), such that the center frequency of each single Voigt profile is given by \(D \pm S/2\). 

\section{Variation of zero-field ODMR spectra from NV centers in different regions of the diamond sample}

\label{sec:variation}

As described in the main text, the ZFS of the NV centers in set A is on average roughly 300 kHz lower than the ZFS of set B NVs near room temperature. 
To investigate whether this may be due to regional variations in strain or the local density of trapped charges, we recorded zero-field ODMR on NV centers from five different regions in the diamond under ambient conditions. In addition to characterizing the shifts in the ZFS, we also looked for splittings of the ODMR line that would similarly indicate local strain or electric fields that could lead to ZFS shifts. The results are displayed in Fig.~\ref{fig:zfs_survey}. 
We see that NV centers in set A (region 1 in Fig.~\ref{fig:zfs_survey}) display a relatively large spread in their zero-field splittings near room temperature relative to centers from other regions. We note that this does not necessarily reflect a real shift in the ZFS. An apparent shift could be caused by the presence of a dark state induced by a combination of local electric/strain fields and microwave polarization \cite{mittiga2018imaging}. Besides the variation in the set A NV centers, we also observe that NV centers in region 5 had consistently larger splittings in their zero-field ODMR lines relative to NV centers in other regions. These observations suggest that there are likely locations within the diamond with higher concentrations of charged defects and/or local strain. This is also consistent with the large variations in the density of native NV$^{-}$ centers throughout the diamond sample. 

\section{Reproduction of prior results and additional comparisons}
\label{sec:additional_comparisons}

In Fig.~4 of the main text we reproduce the models from prior works that have examined the ZFS temperature dependence over at least a 100 K temperature range \cite{chen2011temperature, toyli2012measurement, doherty2014temperature, li2017temperature, barson2019temperature} and compare them to the model presented in this work. An additional work, Ref.~\cite{lourette2022temperature}, measured the ZFS between 75 and 400 K, but did not publish their fit parameters and so was not included in the main text figure. In this section we review these prior models and experimental data sets in more detail. 
We note that two of the prior works do not consider the absolute value of the ZFS, and instead only present the deviation of the ZFS from its zero-temperature limit \cite{doherty2014temperature, barson2019temperature}. In order to compare the data and models from these works to other absolute ZFS results, we add a constant offset equal to the zero-temperature limit ZFS found in this work, \(D_{0}=2877.38(3)\) MHz. 

\subsection{Reproduction of prior models}
\label{sec:model_reproduction}

The fit from each prior work is reproduced according to the model function and fit parameters published in that work. The fit functions and original fit parameters are presented in Table~\ref{tab:comp-models}. The fit functions from Refs.~\cite{chen2011temperature} and \cite{toyli2012measurement} are purely empirical fifth- and third-order polynomials respectively. Doherty et al.~motivate some parts of their model with arguments similar to those invoked in this work and in Ref.~\cite{tang2022first}, observing that the ZFS can be modeled by summing contributions from thermal expansion and electron-phonon interactions \cite{doherty2014temperature}. They calculate the thermal expansion contribution directly using prior results describing the diamond coefficient of thermal expansion \cite{sato2002thermal} and the ZFS response to hydrostatic pressure \cite{doherty2014electronic}. Doherty et al.~argue that electron-phonon interactions contribute only starting at fourth order in temperature such that the final ZFS temperature dependence expression from Doherty et al. is a fifth-order polynomial with two free parameters that describe the electron-phonon contributions. The expression does not include a constant term or linear terms and only models the deviation of the ZFS from the zero-temperature limit. Li et al.~present a fit based on a modified Varshni equation \cite{li2017temperature}. The equation transitions from quartic behavior at low temperatures to quadratic behavior at higher temperatures and so is similar to a polynomial in practice. The model presented by Barson et al.~\cite{barson2019temperature} builds on that developed in Ref.~\cite{doherty2014temperature}. Instead of using a power series to model the coefficient of thermal expansion, Barson et al.~use the expression from Ref.~\cite{reeber1996thermal}, which consists of a weighted sum of Einstein heat capacities as in Eq.~8 of the main text of this work. Barson et al.~additionally add a sixth-order term to the polynomial, resulting in a final expression with three free parameters. They fit their model to an artificial data set covering 0-700 K, which is generated from prior works \cite{doherty2014temperature, toyli2012measurement}. The data sets from these two prior works feature a discontinuity at room temperature, which is manually removed, and the final artificial data set only describes the deviation of the ZFS from the zero-temperature limit. Finally, Lourette et al.~state that they fit their data using a fourth-order polynomial, but do not provide the extracted fit parameters in their manuscript, and so we do not attempt to reproduce their fit here \cite{lourette2022temperature}.

\subsection{Comparison to prior experimental data}

\begin{figure}[b]
\includegraphics[width=0.6\textwidth]{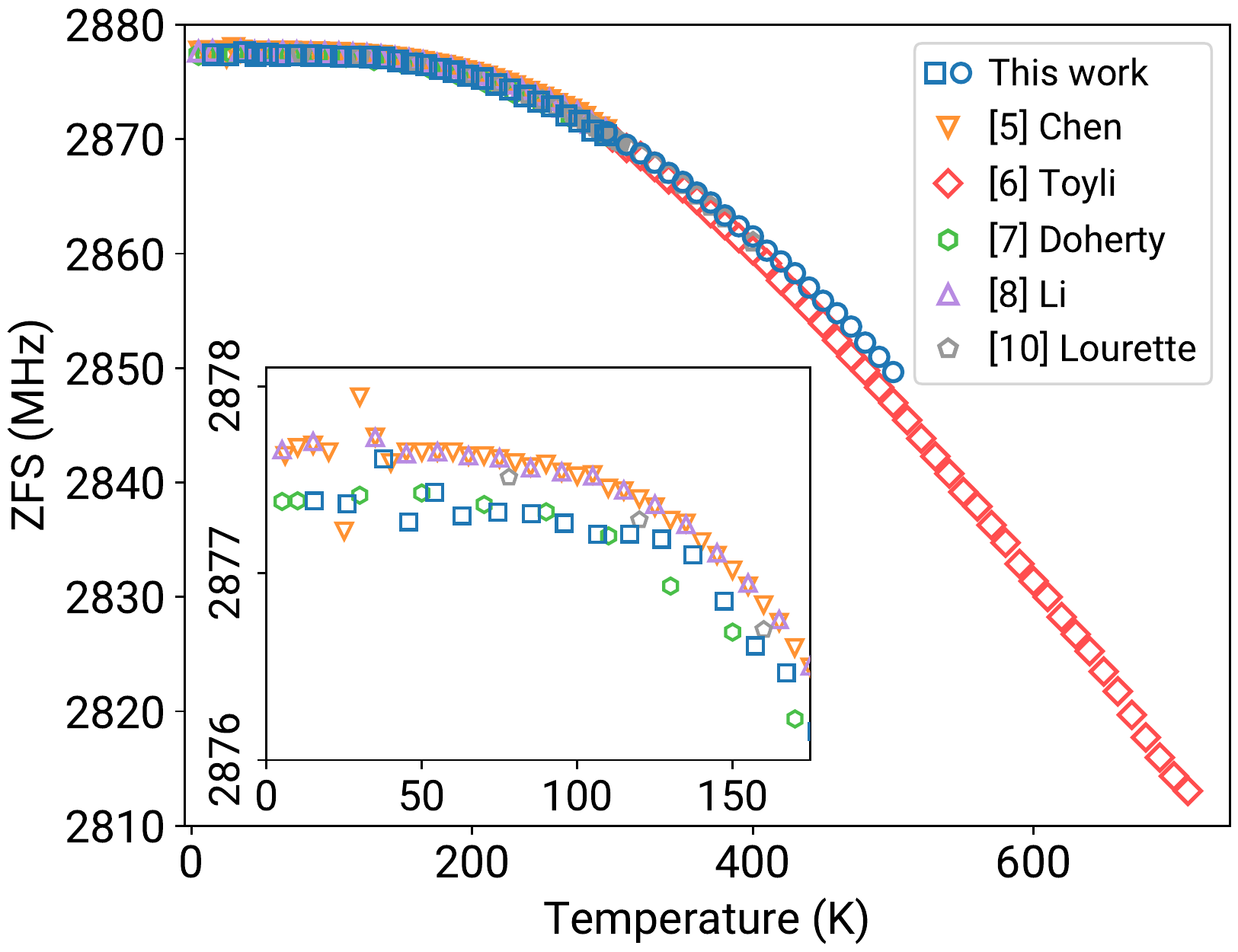}
\caption{\label{fig:comp-data}
Comparison to prior experimental measurements of the ZFS temperature dependence. Prior experimental data is digitized and reproduced from Refs. \cite{toyli2012measurement, chen2011temperature, doherty2014temperature, li2017temperature, lourette2022temperature}. Inset: Low-temperature behavior of data.
}
\end{figure}

Fig.~\ref{fig:comp-data} displays the experimental data collected for this work alongside data from prior works \cite{chen2011temperature, toyli2012measurement, doherty2014temperature, li2017temperature, lourette2022temperature}. The experimental data was digitized from Fig.~3a in Ref.~\cite{chen2011temperature}, Fig.~5c in Ref.~\cite{toyli2012measurement}, Fig.~2a in Ref.~\cite{doherty2014temperature}, Fig.~1b in Ref.~\cite{li2017temperature}, and Fig.~3e (black data points for \(^{14}\)NV) in Ref.~\cite{lourette2022temperature}. The data was digitized using WebPlotDigitizer \cite{Rohatgi2022}. Data from Refs.~\cite{chen2011temperature, doherty2014temperature, li2017temperature, lourette2022temperature} is for NV ensembles while data from Ref.~\cite{toyli2012measurement} is for a single NV center. We note that Li et al.~included  additional data sets for three single NVs each from the same sample which we do not reproduce in this work \cite{li2017temperature}. These single NV data sets are relatively noisy and exhibit ZFS values that are unusually high, lying roughly 1 MHz above the values reported by Li et al.~for their ensemble measurements. Li et al.~speculate that this may be due to impurities in the diamond sample used for single NV measurements \cite{li2017temperature}. Lourette et al.~report additional ZFS measurements of NV ensembles in \(^{15}\)N-enriched diamond samples \cite{lourette2022temperature}. We do not reproduce these results here, but note that Lourette et al.~observe a small ZFS isotope shift of about +120 kHz in \(^{15}\)NV centers.

The available prior data sets of the NV center ZFS temperature dependence disagree in several ways, as seen in Fig~\ref{fig:comp-data}. Below room temperature, the data and fits from Refs.~\cite{chen2011temperature} and \cite{li2017temperature} exhibit very similar ZFS values, which are approximately 300 kHz above the ZFS values measured in this work. It is likely that this difference is a sample dependent effect, as the temperature dependence of the ZFS is negligible between 0 and 50 K and the difference between our measurements and the measurements from Refs~\cite{chen2011temperature} and \cite{li2017temperature} is consistent up to room temperature. The variation of the ZFS described in Sec.~\ref{sec:variation} provides additional evidence that this is a plausible explanation. Near room temperature, we observe that the ZFS of set B NVs (which were used for high temperature measurements) is similarly about 300 kHz above the ZFS of set A NVs (which were used for low temperature measurements). The set B NV measurements are consistent with other prior measurements near room temperature. 

At higher temperatures, we also observe variations between the different works in the measured slopes of the ZFS temperature dependence. While near room temperature our measurements (for set B NVs) and the measurements from Refs.~\cite{toyli2012measurement} and \cite{lourette2022temperature} agree to within 100 kHz, by 400 K, all three measurements differ significantly. Lourette et al.'s measurement is approximately 500 kHz below the measurement made for this work \cite{lourette2022temperature}, and Toyli et al.'s measurement is approximately 750 kHz below Lourette et al.'s measurement. Toyli et al.'s measurement is approximately 1.25 MHz below our measurement \cite{toyli2012measurement} at the same temperature. We suggest that these differences are likely due to systematic errors in measuring the temperature of the diamond at high temperatures. 

\begin{figure}[b]
\includegraphics[width=0.98\textwidth]{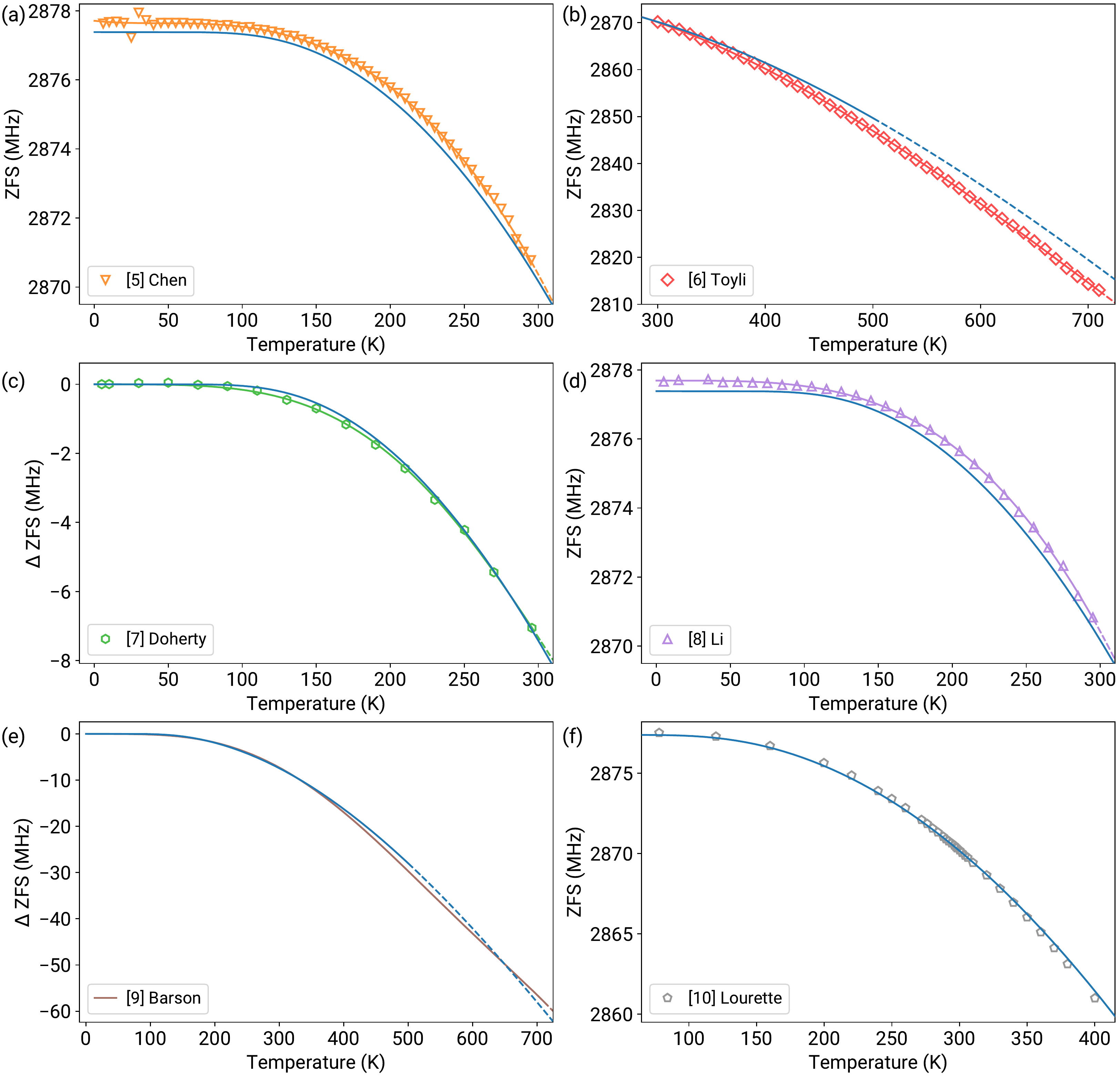}
\caption{\label{fig:comp-sep}
Pairwise comparisons between this work and a given prior work. In each panel, the model and data set from a prior work is plotted alongside the model from this work (solid and dashed blue line). Note that the model presented by Barson et al. in Ref.~\cite{barson2019temperature} is fit to artificial data generated from Refs.~\cite{doherty2014temperature, toyli2012measurement}. As such, there is no experimental data shown in panel (e). Lourette et al.~do not provide the fit parameters for the fourth-order polynomial they use to fit to their data set so only the data from Ref.~\cite{lourette2022temperature} is shown in panel (f). For panels (c) and (e), in accordance with what was published in Refs.~\cite{doherty2014temperature} and \cite{barson2019temperature}, the ZFS deviation from the zero-temperature limit is shown rather than the absolute ZFS. As in the main text, the solid fit lines become dashed outside the range of experimental data used for the fit. For clarity, the experimental data from this work is not replotted. 
}
\end{figure}

\subsection{Pairwise comparisons}

Fig.~4 in the main text and Fig.~\ref{fig:comp-data} compare the prior models and data sets respectively from this prior works on the same plot. Fig.~\ref{fig:comp-sep} presents the same data in a pairwise format so as to allow for easier comparison between this work and a specific prior work. 

\subsection{Fits of prior models to experimental data collected for this work}

\begin{figure}
\includegraphics[width=0.97\textwidth]{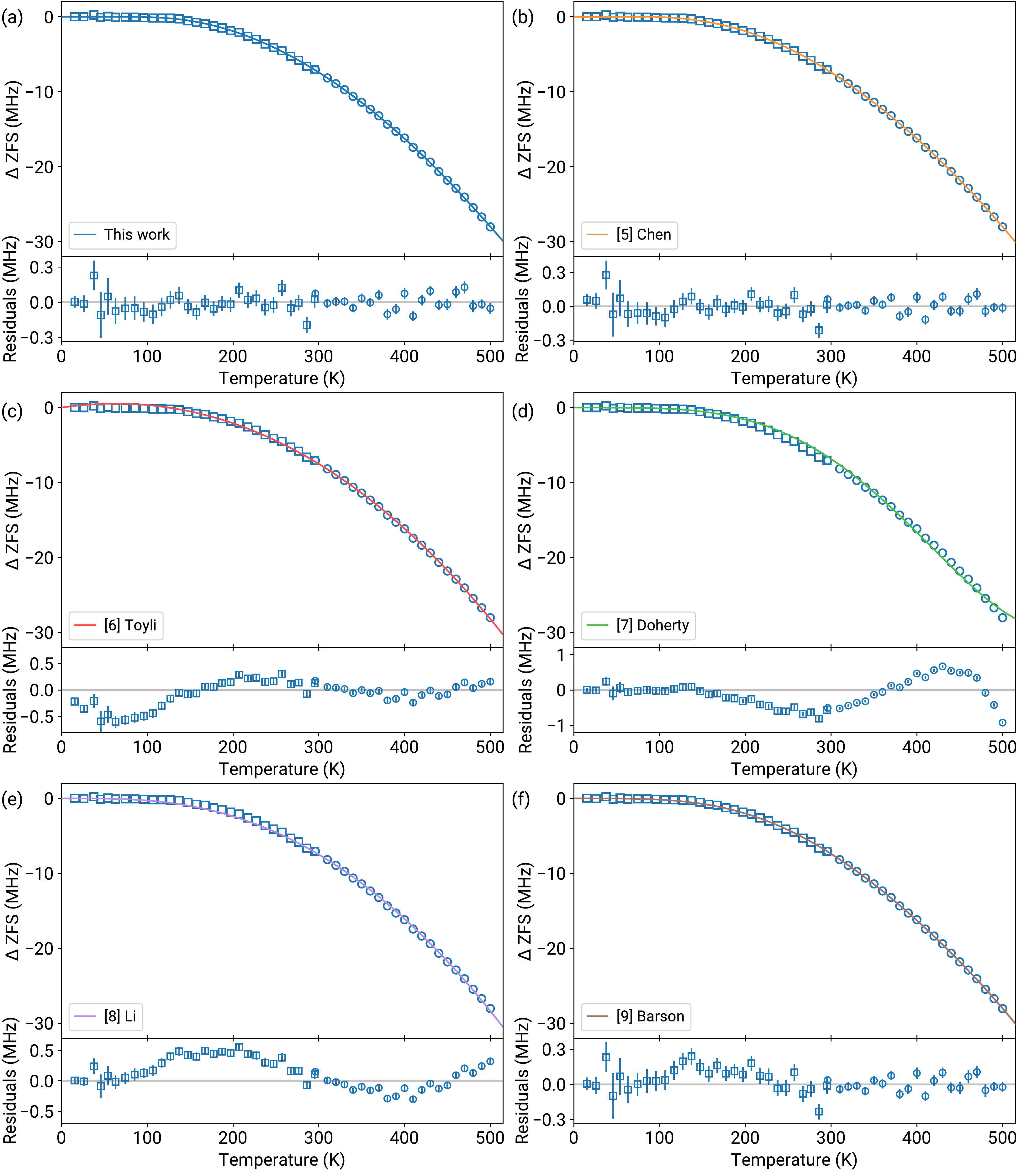}
\caption{\label{fig:prior_models_our_data}
Fits of prior models to ZFS deviation data collected for this work. The models from Refs.~\cite{toyli2012measurement, doherty2014electronic, li2017temperature} clearly stray from the experimental data. The other models provide reasonable fits to the data. The fit parameters and reduced chi squared metrics are shown in Table~\ref{tab:comp-models}.
}
\end{figure}

In this section we compare the quality of the fits provided by the model proposed in this work and the models from prior works, by fitting the models to the same data set. Because goodness-of-fit metrics are strongly affected by the 300 kHz offset between the set A and B measurements, we fit the models to ZFS deviation data generated by subtracting the zero-temperature ZFS limit given in the main text from each of the set A and B measurements performed for this work, and an additional 300 kHz from the set B measurements to get rid of the offset. The resulting fits are presented in Table~\ref{tab:comp-models} and Fig.~\ref{fig:prior_models_our_data}. Our model and the fifth order power series used by Chen et al.~\cite{chen2011temperature} both provide excellent fits to the data, with reduced chi squared metrics \(\chi_{\nu}^{2}\) of 1.50. While the power series from Chen et al. provides a good fit over the temperature range probed in this work, we emphasize that it is not suitable for predicting the ZFS at higher temperatures, and that it would perform worse if used to fit ZFS data covering a broader temperature range, as suggested by the high temperature ZFS measurements made by Toyli et al.~\cite{toyli2012measurement}. The model from Barson et al.~\cite{barson2019temperature} is slightly worse according to its reduced chi squared of \(\chi_{\nu}^{2} = 2.23\). Our model and the models from Refs.~\cite{chen2011temperature, barson2019temperature} provide fits to the data that are visually similar. The models from Refs.~\cite{toyli2012measurement, doherty2014temperature, li2017temperature} provide relatively poor fits with high reduced chi squared metrics and visually apparent departures from the experimental data.

\subsection{Fits of model proposed in this work to experimental data from prior works}

\begin{figure}
\includegraphics[width=0.97\textwidth]{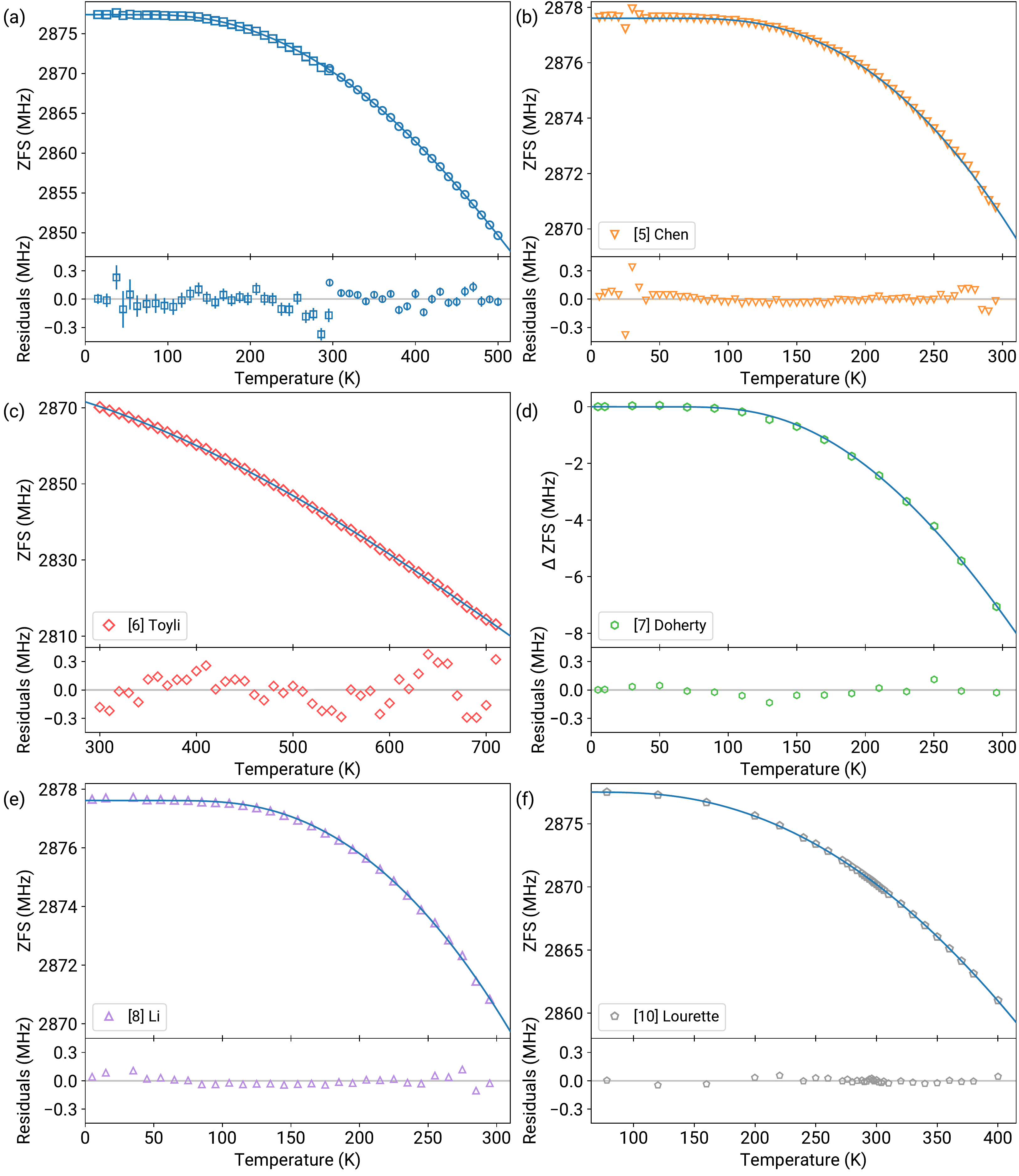}
\caption{\label{fig:our_model_prior_data}
Fits of the model proposed in this work to prior data sets. The energies of the representative phonon modes in the model are fixed to the energies reported in the main text. For fitting to the data from Ref.~\cite{doherty2014temperature} there are two free parameters, the coefficients \(c_{1,2}\) in Eq.~3 of the main text. For fitting to the data from the other references there is an additional free parameter, \(D_{0}\) the zero-temperature ZFS limit. Panel (a) is the same as Fig.~2 of the main text and is reproduced here to allow for easier comparison. Because error bars are not available for the prior data sets, we demonstrate the fit qualities by fixing the range of the residuals shown in the bottom plot of each panel to [-0.45, +0.45] MHz. The residuals across all temperatures and data sets fall within this range, indicating that the model can provide an excellent description of all prior data sets in addition to the data set collected for this work. Note that the accuracy of the prior data sets may be limited by the digitization procedure.
}
\end{figure}

Because of the discrepancies between the ZFS data collected for this work and for prior works, the model and fit parameters described in the main text is not able to describe all the available data simultaneously. Here we demonstrate that a version of the model with fixed representative phonon mode energies can provide an excellent description of all the available data by allowing the coefficients and zero-temperature ZFS limit to be free parameters. We fix the mode energies to the values reported in the main text, \(\Delta_{1}=58.73\) meV and \(\Delta_{2}=145.5\) meV. Fig.~\ref{fig:our_model_prior_data} shows the resulting fits to the original data sets from Refs.~\cite{toyli2012measurement, chen2011temperature, doherty2014temperature, li2017temperature, lourette2022temperature}. Despite the discrepancies between the different data sets, we see that the model with fixed energies does an excellent job of describing each individual data set. This indicates that the success of the model is not specific to the data set collected for this work, and likewise that the representative phonon mode energies reported in the main text are not specific to any given NV or sample. 

\begin{table*}
    {\scriptsize
    \begin{tabularx}{0.98\textwidth}{|>{\arraybackslash}p{2.1cm} >{\arraybackslash}p{4.5cm} >{\arraybackslash}X >{\arraybackslash}X | >{\arraybackslash}p{3.0cm} >{\arraybackslash}p{1.0cm} |} 
    \hline
    Reference & Functional form & Fit free \newline parameters & Original fit & Refit to ZFS deviation data from this work & Refit \(\chi_{\nu}^{2}\) \\
    \hline
    This work & \(D(T) = D_{0} + c_{1}n_{1} + c_{2}n_{2}\) & \(D_{0}\) (MHz) & 2877.38(3) & n/a & 1.50 \\
    & & \(c_{1}\) (MHz) & -55(7) & \num{-75(6)} & \\
    & & \(\Delta_{1}\) (meV) & 59(2) & 63.9(15) & \\
    & & \(c_{2}\) (MHz)  & \num{-2.5(19)e2} & \num{-2.5(3)e2} & \\
    & & \(\Delta_{2}\) (meV) & 146(8) & 163(12) & \\
    \hline
    \cite{chen2011temperature} Chen et al.~& \(D(T)=\sum_{i=0}^{5}d_{i}T^{i}\) & \(d_{0}\) (MHz) & 2877.71 & n/a & 1.50 \\
    & & \(d_{1}\) (MHz/K) & \num{-4.625e-3} & \num{-5.6(14)e-3} & \\
    & & \(d_{2}\) (MHz/K\(^{2}\)) & \num{1.067e-4} & \num{1.6(2)e-4} & \\
    & &  \(d_{3}\) (MHz/K\(^{3}\)) & \num{-9.325e-7} & \num{-1.31(11)e-6} & \\
    & & \(d_{4}\) (MHz/K\(^{4}\)) & \num{1.739e-9} & \num{2.3(2)e-9} & \\
    & & \(d_{5}\) (MHz/K\(^{5}\)) & \num{-1.838e-12} & \num{-1.48(18)e-12} & \\
    \hline
    \cite{toyli2012measurement} Toyli et al.~& \(D(T)=\sum_{i=0}^{3}a_{i}T^{i}\) & \(a_{0}\) (MHz) & 2869.7(9) & n/a & 9.90 \\
    & &\(a_{1}\) (MHz/K) & \num{9.7(6)e-2} & \num{1.65(3)e-2} & \\
    & & \(a_{2}\) (MHz/K\(^{2}\)) & \num{-3.7(1)e-4} & \num{-1.274(18)e-4} & \\
    & & \(a_{3}\) (MHz/K\(^{3}\)) & \num{1.7(1)e-7} & \num{-3.7(3)e-8} & \\
    \hline
    \cite{doherty2014temperature} Doherty et al.~& \(\Delta D(T)=-AB\sum_{i=1}^{5}(e_{i-1}/i) T^{i}\) & \(b_{4}\) (MHz/K\(^{4}\)) & \num{18.7(4)e-10} & \num{1.161(3)e-9} & 61.07 \\
    & \hspace{1.3cm}\(+ \sum_{i=4}^{5}b_{i}T^{i}\) & \(b_{5}\) (MHz/K\(^{5}\)) & -\num{41(2)e-13} & \num{-1.932(6)e-12} & \\
    \hline
    \cite{li2017temperature} Li et al.~& \(D(T)=D_{0}-[AT^{4}/(T+B)^{2}]\) & \(D_{0}\) (MHz) & 2877.69(1) & n/a & 19.86 \\
    &  & \(A\) (MHz/K\(^{2}\)) & \num{5.6(4)e3} & \num{2.029(10)e-4} & \\
    &  & \(B\) (K) & 490(32) & \num{168.8(15)} & \\
    \hline
    \cite{barson2019temperature} Barson et al.~& \(\Delta D(T) = -AB(\Delta V / V) + \sum_{i=4}^{6}b_{i}T^{i}\) & \(b_{4}\) (MHz/K\(^{4}\)) & \num{-1.44(8)e-9} & \num{-1.813(18)e-9} & 2.23 \\
    &  & \(b_{5}\) (MHz/K\(^{5}\)) & \num{3.1(3)e-12} & \num{4.90(9)e-12} & \\
    &  & \(b_{6}\) (MHz/K\(^{6}\)) & \num{-1.8(3)e-15} & \num{-3.79(10)e-15} & \\
    \hline
    \end{tabularx}
    \caption{Comparison of the functional forms and fit parameters of the available models of the ZFS temperature dependence. Note that some works originally report parameters in units of GHZ rather than MHz. For consistency, we have converted to MHz where this is the case. Here we match the notation of the functional forms used in the original works. The first terms in the expressions from Refs.~\cite{doherty2014temperature} and \cite{barson2019temperature} describe the effect of thermal expansion and are not free parameters of the fit. In these works the effect of thermal expansion is expressed in terms of the NV hydrostatic pressure shift \(A\), the diamond bulk modulus \(B\), and the diamond coefficient of thermal expansion. 
    Doherty et al. use values of \(A=14.6\) MHz/GPa \cite{doherty2014electronic} and \(B = 442\) GPa for the hydrostatic pressure shift and bulk modulus respectively. The coefficient of thermal expansion terms \(e_{i}\) are derived from Ref.~\cite{sato2002thermal} and are given by Doherty et al. as \(ABe_{1}/2=\num{39.7e-7}\) MHz / T\(^{2}\), \(ABe_{2}/3=\num{-91.6e-9}\) MHz / T\(^{3}\), \(ABe_{3}/4=\num{70.6e-11}\) MHz / T\(^{4}\), \(ABe_{4}/5=\num{-60.0e-14}\) MHz / T\(^{5}\). Barson et al. use the same values as Doherty et al. for the hydrostatic pressure shift \(A\) and the bulk modulus \(B\), and describe the coefficient of thermal expansion \(\Delta V / V\) using the expression from Ref.~\cite{reeber1996thermal}, which is a weighted sum over Einstein heat capacities and is not given by Barson et al. in their manuscript. In contrast, in Ref.~\cite{li2017temperature} \(A\) and \(B\) refer to free parameters in the fit. The last two columns in the table describe the fits of the models to the ZFS deviation data collected for this work, as shown in Fig.~\ref{fig:prior_models_our_data}.}
    \label{tab:comp-models}
    }
\end{table*}

\section{Comparison of phonon energies identified in this work and prior works}

In this work we describe the temperature dependence of the NV center ZFS by considering the effects of two representative phonon modes. Several prior works have used a similar technique to describe other properties of the NV center that are physically rooted in interactions with phonons. In this section we compare the various energies that are reported by these works. In this work, the fit described in the main text yields energies of \(\Delta_{1}=59(2)\) meV and \(\Delta_{2}=146(8)\). Fitting the same model to data of the deviation of the ZFS from its zero-temperature limit (see Section~\ref{sec:model_reproduction}, Fig.~\ref{fig:prior_models_our_data}, and Table~\ref{tab:comp-models} for details) we find slightly higher but consistent energies of 63.9(15) and 163(12) meV. In Ref.~\cite{cambria2022temperature} we used a similar model to describe the temperature dependence of the NV spin-phonon relaxation rates. The energies reported in that work are 68.2(17) and 167(12) meV. Studies of the NV optical spectrum have noted that several prominent features of the phonon sideband are associated with particular phonon energies. Davies' model of the sideband is based on a continuous electron-phonon coupling spectrum with peaks at 69 and 154 meV \cite{davies1974vibronic}. Collins, Stanley, and Woods observe phonon replicas of the zero-phonon line at spacings of 61 and 137 meV \cite{collins1987nitrogen}. In another experimental work, Kehaysias et al. identify significant features in the phonon sideband at 64, 122, 138, 153, and 163 meV \cite{kehayias2013infrared}. Alkauskas et al. study the NV optical spectrum from first principles and identify a dominant 65 meV vibrational resonance, as well as additional modes at 120, 134, 161, and 167 meV that are highly localized to the NV center \cite{alkauskas2014first}. While some of the energies identified in these works may be linked to individual phonon modes that strongly influence NV center dynamics, other energies instead represent the effect of several modes with similar but not identical energies. 
In this latter case we interpret the identified energy as that of a single ``representative'' phonon mode which stands in for the group of several modes. We expect that such representative modes will tend to have a lower energy than the typical energy of the group of modes it represents, as the lower energy modes in the group have larger occupation numbers. This consideration, as well as the fact that the various physical properties (e.g. the ZFS temperature dependence, spin-phonon relaxation rate temperature dependence, the phonon sideband) will be affected differently by the same spectrum of phonon modes, indicates that the energies identified by examining one physical property cannot be neatly compared to the energies identified by examining a different physical property. Nevertheless, as we argue in the main text, the various phonon-influenced properties of diamond and the NV center derive from the same phonon density of states, and so it is reasonable to expect that different studies will identify energies that roughly correspond to peaks (or are somewhat lower than the peaks) in the phonon density of states. This claim is generally true empirically, as can be seen by comparing the energies in the survey reported here against the measured phonon density of states \cite{bosak2005phonon}.

\section{Generation of lattice constant data from Ref.~\cite{jacobson2019thermal}}

In Ref.~\cite{jacobson2019thermal}, Jacobson and Stoupin survey existing experimental studies of the diamond coefficient of thermal expansion and lattice constant. Building off prior work within the quasiharmonic model of thermal expansion, Jacobson and Stoupin fit the various measurements of the coefficient of thermal expansion to a weighted sum of Einstein heat capacities at four different frequencies. They present two different fits using the same model where Fit 1 (Fit 2) covers experimental data up to 1000 K (2000 K). While both fits provide an excellent description of the experimental up to 1000 K, Fit 2 yields activation energies above the energy of the debye temperature, inconsistent with the quasiharmonic model. Jacobson and Stoupin further accurately reproduce measurements of the lattice constant up to 1000 K by integrating both fits of the coefficient of thermal expansion according to Eq.~9 from the main text with \(a_{0}=3.56712\) \si{\angstrom} at \(T_{0} = 298\) K \cite{stoupin2010thermal}. NV center measurements have been demonstrated up to a maximum temperature of 1000 K \cite{liu2019coherent}, and so for Fig.~3a of the main text we represent the diamond lattice constant using the curve calculated from Fit 1 of Ref.~\cite{jacobson2019thermal}. The fit parameters we find by fitting Eq.~11 of the main text to this generated lattice constant data are \(a_{0}=3.56677 \si{\angstrom}\), \(b_{1}=\num{2.027e-03} \si{\angstrom}\) and \(b_{2}=\num{2.609e-02} \si{\angstrom}\).

\section{Complete data set collected for this work}
The complete set of experimental data collected for this work is presented in Table~\ref{tab:complete_data}.

\begin{sidewaystable*}[b]
    {\scriptsize
    \begin{tabularx}{0.9\textwidth}{c r l || c r l || c r l || c r l || c r l } \\
    \hline
    \multicolumn{1}{c}{NV} & \multicolumn{1}{c}{\(T\) (K)} & \multicolumn{1}{c}{\(D\) (MHz)} & \multicolumn{1}{c}{NV} & \multicolumn{1}{c}{\(T\) (K)} & \multicolumn{1}{c}{\(D\) (MHz)} & \multicolumn{1}{c}{NV} & \multicolumn{1}{c}{\(T\) (K)} & \multicolumn{1}{c}{\(D\) (MHz)} & \multicolumn{1}{c}{NV} & \multicolumn{1}{c}{\(T\) (K)} & \multicolumn{1}{c}{\(D\) (MHz)} & \multicolumn{1}{c}{NV} & \multicolumn{1}{c}{\(T\) (K)} & \multicolumn{1}{c}{\(D\) (MHz)} \\
    \hline
    A1 & 15.3 & 2877.36(12) & A2 & 15.3 & 2877.23(12) & A3 & 15.3 & 2877.52(10) & A4 & 15.3 & 2877.41(12) & A5 & 15.3 & 2877.21(14) \\A1 & 26.0 & 2877.63(16) & A2 & 26.0 & 2877.44(15) & A3 & 26.0 & 2877.41(13) & A4 & 26.0 & 2877.27(16) & A5 & 26.0 & 2877.0(2) \\A1 & 37.7 & 2877.5(3) & A2 & 37.7 & 2877.4(3) & A3 & 37.7 & 2877.0(3) & A4 & 37.7 & 2878.1(2) & A5 & 37.7 & 2877.9(4) \\A1 & 45.8 & 2876.5(5) & A2 & 45.8 & 2877.9(4) & A3 & 45.8 & 2877.0(4) & A4 & 45.8 & 2877.6(4) & A5 & 45.8 & 2877.3(4) \\A1 & 54.0 & 2877.1(4) & A2 & 54.0 & 2877.5(5) & A3 & 54.1 & 2877.6(3) & A4 & 54.1 & 2877.3(4) & A5 & 54.2 & 2877.4(3) \\A1 & 62.9 & 2877.6(3) & A2 & 62.9 & 2877.4(3) & A3 & 63.0 & 2876.9(2) & A4 & 63.0 & 2877.4(2) & A5 & 63.0 & 2877.4(3) \\A1 & 74.4 & 2877.5(3) & A2 & 74.4 & 2877.0(3) & A3 & 74.4 & 2877.3(2) & A4 & 74.4 & 2877.29(18) & A5 & 74.4 & 2877.2(2) \\A1 & 85.2 & 2877.0(3) & A2 & 85.2 & 2877.8(3) & A3 & 85.2 & 2877.2(2) & A4 & 85.2 & 2877.31(18) & A5 & 85.2 & 2877.4(3) \\A1 & 95.9 & 2877.35(18) & A2 & 95.9 & 2877.3(2) & A3 & 95.9 & 2877.27(20) & A4 & 95.8 & 2877.19(17) & A5 & 95.9 & 2877.20(18) \\A1 & 106.5 & 2877.1(2) & A2 & 106.5 & 2877.4(2) & A3 & 106.5 & 2877.36(18) & A4 & 106.5 & 2877.17(17) & A5 & 106.5 & 2877.07(17) \\A1 & 116.9 & 2877.6(2) & A2 & 116.9 & 2877.3(2) & A3 & 116.9 & 2877.03(16) & A4 & 116.9 & 2877.07(17) & A5 & 116.9 & 2877.24(18) \\A1 & 127.1 & 2876.97(19) & A2 & 127.1 & 2877.3(2) & A3 & 127.1 & 2877.57(16) & A4 & 127.1 & 2876.95(15) & A5 & 127.0 & 2877.15(15) \\A1 & 137.3 & 2877.44(17) & A2 & 137.2 & 2876.93(17) & A3 & 137.3 & 2877.11(15) & A4 & 137.3 & 2876.87(15) & A5 & 137.3 & 2877.18(17) \\A1 & 147.3 & 2876.71(15) & A2 & 147.3 & 2876.99(16) & A3 & 147.3 & 2876.60(15) & A4 & 147.3 & 2876.80(14) & A5 & 147.3 & 2877.13(15) \\A1 & 157.3 & 2876.61(18) & A2 & 157.3 & 2876.64(17) & A3 & 157.3 & 2876.47(14) & A4 & 157.4 & 2876.66(14) & A5 & 157.4 & 2876.55(14) \\A1 & 167.3 & 2876.54(15) & A2 & 167.3 & 2876.51(18) & A3 & 167.3 & 2876.59(17) & A4 & 167.4 & 2876.15(13) & A5 & 167.3 & 2876.53(13) \\A1 & 177.3 & 2876.27(15) & A2 & 177.3 & 2876.16(17) & A3 & 177.3 & 2876.01(15) & A4 & 177.3 & 2876.22(13) & A5 & 177.3 & 2875.97(13) \\A1 & 187.2 & 2876.13(14) & A2 & 187.2 & 2876.07(17) & A3 & 187.2 & 2875.96(13) & A4 & 187.2 & 2875.56(12) & A5 & 187.2 & 2875.83(14) \\A1 & 197.1 & 2875.49(18) & A2 & 197.1 & 2876.01(17) & A3 & 197.1 & 2875.53(15) & A4 & 197.1 & 2875.38(12) & A5 & 197.1 & 2875.37(13) \\A1 & 207.2 & 2875.12(17) & A2 & 207.1 & 2875.49(16) & A3 & 207.2 & 2875.48(14) & A4 & 207.2 & 2875.21(13) & A5 & 207.1 & 2875.06(13) \\A1 & 217.2 & 2874.90(16) & A2 & 217.2 & 2875.14(18) & A3 & 217.2 & 2874.85(14) & A4 & 217.2 & 2874.67(13) & A5 & 217.2 & 2874.35(16) \\A1 & 227.1 & 2874.64(19) & A2 & 227.1 & 2874.56(17) & A3 & 227.1 & 2874.34(15) & A4 & 227.1 & 2874.26(13) & A5 & 227.1 & 2874.07(15) \\A1 & 237.6 & 2873.72(19) & A2 & 237.6 & 2874.0(2) & A3 & 237.6 & 2873.89(14) & A4 & 237.6 & 2873.69(12) & A5 & 237.6 & 2873.65(15) \\A1 & 247.3 & 2873.25(18) & A2 & 247.2 & 2873.52(19) & A3 & 247.2 & 2873.22(14) & A4 & 247.2 & 2873.29(14) & A5 & 247.2 & 2873.21(13) \\A1 & 257.1 & 2872.81(20) & A2 & 257.0 & 2873.11(18) & A3 & 257.1 & 2872.98(14) & A4 & 257.1 & 2872.75(13) & A5 & 257.0 & 2872.87(17) \\A1 & 267.3 & 2872.36(15) & A2 & 267.2 & 2872.02(18) & A3 & 267.2 & 2871.90(14) & A4 & 267.2 & 2871.87(15) & A5 & 267.3 & 2872.18(14) \\A1 & 276.4 & 2871.77(15) & A2 & 276.4 & 2871.80(18) & A3 & 276.4 & 2871.44(15) & A4 & 276.4 & 2871.45(14) & A5 & 276.4 & 2871.50(14) \\A1 & 285.9 & 2870.77(14) & A2 & 285.9 & 2870.98(17) & A3 & 285.9 & 2870.72(15) & A4 & 286.0 & 2870.74(14) & A5 & 286.0 & 2870.74(14) \\A1 & 295.1 & 2870.77(19) & A2 & 295.2 & 2870.5(2) & A3 & 295.2 & 2870.40(14) & A4 & 295.1 & 2870.20(13) & A5 & 295.1 & 2870.17(15) \\B1 & 296.0 & 2870.53(8) & B2 & 296.0 & 2870.67(8) & B3 & 296.0 & 2870.74(7) & B4 & 296.0 & 2870.57(9) & B5 & 296.0 & 2870.72(8) \\B1 & 310.0 & 2869.39(8) & B2 & 310.0 & 2869.48(8) & B3 & 310.0 & 2869.58(7) & B4 & 310.0 & 2869.64(11) & B5 & 310.0 & 2869.64(10) \\B1 & 320.0 & 2868.75(9) & B2 & 320.0 & 2868.66(9) & B3 & 320.0 & 2868.80(8) & B4 & 320.0 & 2868.65(11) & B5 & 320.0 & 2868.84(9) \\B1 & 330.0 & 2867.96(7) & B2 & 330.0 & 2867.67(8) & B3 & 330.0 & 2867.93(7) & B4 & 330.0 & 2868.09(9) & B5 & 330.0 & 2868.16(8) \\B1 & 340.0 & 2867.18(7) & B2 & 340.0 & 2867.06(8) & B3 & 340.0 & 2867.07(8) & B4 & 340.0 & 2866.98(9) & B5 & 340.0 & 2867.03(8) \\B1 & 350.0 & 2866.24(8) & B2 & 350.0 & 2866.15(9) & B3 & 350.0 & 2866.38(9) & B4 & 350.0 & 2866.28(11) & B5 & 350.0 & 2866.41(10) \\B1 & 360.0 & 2865.24(8) & B2 & 360.0 & 2865.26(9) & B3 & 360.0 & 2865.37(8) & B4 & 360.0 & 2865.50(11) & B5 & 360.0 & 2865.44(10) \\B1 & 370.0 & 2864.61(8) & B2 & 370.0 & 2864.43(8) & B3 & 370.0 & 2864.51(8) & B4 & 370.0 & 2864.58(11) & B5 & 370.0 & 2864.31(11) \\B1 & 380.0 & 2863.56(9) & B2 & 380.0 & 2863.23(9) & B3 & 380.0 & 2863.14(8) & B4 & 380.0 & 2863.56(14) & B5 & 380.0 & 2863.53(11) \\B1 & 390.0 & 2862.40(9) & B2 & 390.0 & 2862.31(10) & B3 & 390.0 & 2862.57(8) & B4 & 390.0 & 2862.24(12) & B5 & 390.0 & 2862.48(12) \\B1 & 400.0 & 2861.63(9) & B2 & 400.0 & 2861.50(10) & B3 & 400.0 & 2861.31(9) & B4 & 400.0 & 2861.37(15) & B5 & 400.0 & 2861.60(14) \\B1 & 410.0 & 2860.17(9) & B2 & 410.0 & 2860.34(11) & B3 & 410.0 & 2860.34(10) & B4 & 410.0 & 2860.12(9) & B5 & 410.0 & 2860.38(9) \\B1 & 420.0 & 2859.46(10) & B2 & 420.0 & 2859.10(11) & B3 & 420.0 & 2859.42(10) & B4 & 420.0 & 2859.41(11) & B5 & 420.0 & 2859.33(9) \\B1 & 430.0 & 2858.30(9) & B2 & 430.0 & 2858.17(11) & B3 & 430.0 & 2858.27(9) & B4 & 430.0 & 2858.40(12) & B5 & 430.0 & 2858.32(9) \\B1 & 440.0 & 2856.93(9) & B2 & 440.0 & 2856.89(9) & B3 & 440.0 & 2857.25(10) & B4 & 440.0 & 2857.05(11) & B5 & 440.0 & 2856.96(9) \\B1 & 450.0 & 2855.78(10) & B2 & 450.0 & 2855.78(11) & B3 & 450.0 & 2855.93(10) & B4 & 450.0 & 2856.22(12) & B5 & 450.0 & 2856.02(11) \\B1 & 460.0 & 2855.03(11) & B2 & 460.0 & 2854.72(12) & B3 & 460.0 & 2854.89(11) & B4 & 460.0 & 2854.76(14) & B5 & 460.0 & 2854.59(12) \\B1 & 470.0 & 2853.75(11) & B2 & 470.0 & 2853.86(12) & B3 & 470.0 & 2853.41(10) & B4 & 470.0 & 2853.64(17) & B5 & 470.0 & 2853.55(12) \\B1 & 480.0 & 2852.14(12) & B2 & 480.0 & 2852.07(13) & B3 & 480.0 & 2852.19(10) & B4 & 480.0 & 2852.59(14) & B5 & 480.0 & 2852.23(12) \\B1 & 490.0 & 2851.12(7) & B2 & 490.0 & 2850.94(9) & B3 & 490.0 & 2850.95(7) & B4 & 490.0 & 2850.82(13) & B5 & 490.0 & 2850.89(13) \\B1 & 500.0 & 2849.93(9) & B2 & 500.0 & 2849.34(10) & B3 & 500.0 & 2849.67(8) & B4 & 500.0 & 2849.73(12) & B5 & 500.0 & 2849.62(14) \\

    \hline
    \end{tabularx}
    \caption{Complete set of experimental measurements of zero-field splitting \(D\) versus temperature \(T\) collected for this work. NVs are indexed first by group and then numerically to distinguish single NVs.}
    }
    \label{tab:complete_data}
\end{sidewaystable*}

\clearpage
\bibliography{SuppReferences.bib}